\documentclass{jfm}
\usepackage{graphicx}
\usepackage{epstopdf, epsfig}
\usepackage{amsmath}
\usepackage{amssymb}
\usepackage{float}
\usepackage{subcaption}

\shorttitle{Study of Asymmetric Cap-Shock Mach Reflection}
\shortauthor{J.K.J. Hew, R. Boswell, C. Federrath, R. Gopalapillai, V. Paramanantham}

\title{Analytical and Numerical Studies of the Non-uniformity induced Type-II Asymmetric Cap Shock Mach Reflection in Over-expanded Supersonic Jets}

\author{Justin Kin Jun Hew\aff{1,2}
  \corresp{\email{u7322062@anu.edu.au}},
  Rod W. Boswell\aff{1}, Christoph Federrath\aff{2}, Rajesh Gopalapillai\aff{3} \and Vinoth Paramanantham\aff{3}}

\affiliation{\aff{1}Space Plasma Power and Propulsion (SP3) Laboratory, Department of Nuclear Physics and Accelerator Applications, Research School of Physics, Australian National University, Canberra, ACT, 2601, Australia
\aff{2}Research School of Astronomy \& Astrophysics, Australian National University, Canberra, ACT, 2611, Australia
\aff{3}Department of Aerospace Engineering, Indian Institute of Technology Madras, Chennai 600036, India}
\begin{document}

\maketitle

\begin{abstract}
A combined analytical and numerical study is conducted to investigate the asymmetric cap-shock non-uniform Mach Reflection (csMR) phenomenon outside of an over-expanded supersonic jet. Prior analytical works have only considered the wedge-induced steady symmetric and asymmetric Mach reflection (MR) configurations, as well as symmetric MR in an open jet. However, there is another structure occurring in nozzle flow fields known as a non-uniformity induced cap-shock pattern. We derive a new analytical model to predict the wave structure of the asymmetric csMR in the absence of internal shocks by extending on a prior symmetric Mach reflection model. Different from the wedge flow case, flow non-uniformity is incorporated by assuming different upstream Mach numbers in both upper and lower domains, where the asymmetry is predicted through averaged flowfields and slipstream inclination angles. The numerical approach utilises an Euler solver for comparisons to the developed theory. It is found that the model adequately predicts the shock structure obtained from numerical simulations, and can be utilised for various sets of parameters to capture the overall direct Mach reflection o[DiMR+DiMR] configuration. The von Neumann criterion is also well captured by the new analytic model, along with the Mach stem profile and shock curvatures. Based on the analytical and numerical observations, a hypothesis is also made regarding the stability of MR structures within an over-expanded jet.

\end{abstract}


\section{Introduction}
Shock reflection phenomena are ubiquitous within virtually all aerospace technologies such as supersonic inlets \citep{babinsky2008sbli}, scramjet engines \citep*{ogawa2012physical}, Busemann intakes \citep*{molder1966busemann}, supersonic/hypersonic air-breathing engines \citep{hunt2019origin}, over-expanded propulsive nozzles \citep{Nasuti2009,Hadjadj2009}, supersonic ejectors \citep{kumar2017shock}, cavity flows \citep{karthick2021shock}, and many more. 

Shock reflections typically result in significant performance degradation such as stagnation pressure losses, broadband shock-associated noise \citep{raman1999supersonic,edgington2019aeroacoustic} and even devastating events like engine unstart and structural damages to the vehicles in question \citep{nave}. Other areas where shock reflection is prevalent is in astrophysics \citep{kevlahan2009shock,foster2010mach, pudritz2013shock,hansen2015numerical,hartigan2016shock}. Colliding curved shock waves can occur within the interstellar medium (ISM), which naturally result in flow non-uniformities and baroclinic generation of vorticity. The instabilities, in particular the associated turbulent flows that develop in these shocked media, are believed to play a key role in star formation \citep*{federrath2008density,molina2012density,federrath2016magnetic}.

Shock reflections are also utilised in inertial confinement fusion (ICF) and other high-energy density applications, where colliding shock waves can induce Richtmyer-Meshkhov instabilities (RMI) due to the misalignment of density fields, which have been proposed to result in performance degradation in nuclear reactors \citep{goldman1999shock}. 

Despite the widespread appearance of shock reflection structures and their associated adverse flow structure interactions, the problem still remains largely unsolved. While there exists a considerable amount of works investigating the low-frequency unsteadiness and dynamical instabilities associated with these interactions, which are highlighted in numerous review articles such as \citet{dolling2001fifty} and \citet{gaitonde2015progress}, 
analytical considerations of the fundamental structure of shock reflection in the near flow field are still lacking, hence the object of the investigation. 

\begin{figure}
\centering
\includegraphics[trim = {0cm, 8cm, 8cm, 5cm}, width = \linewidth]{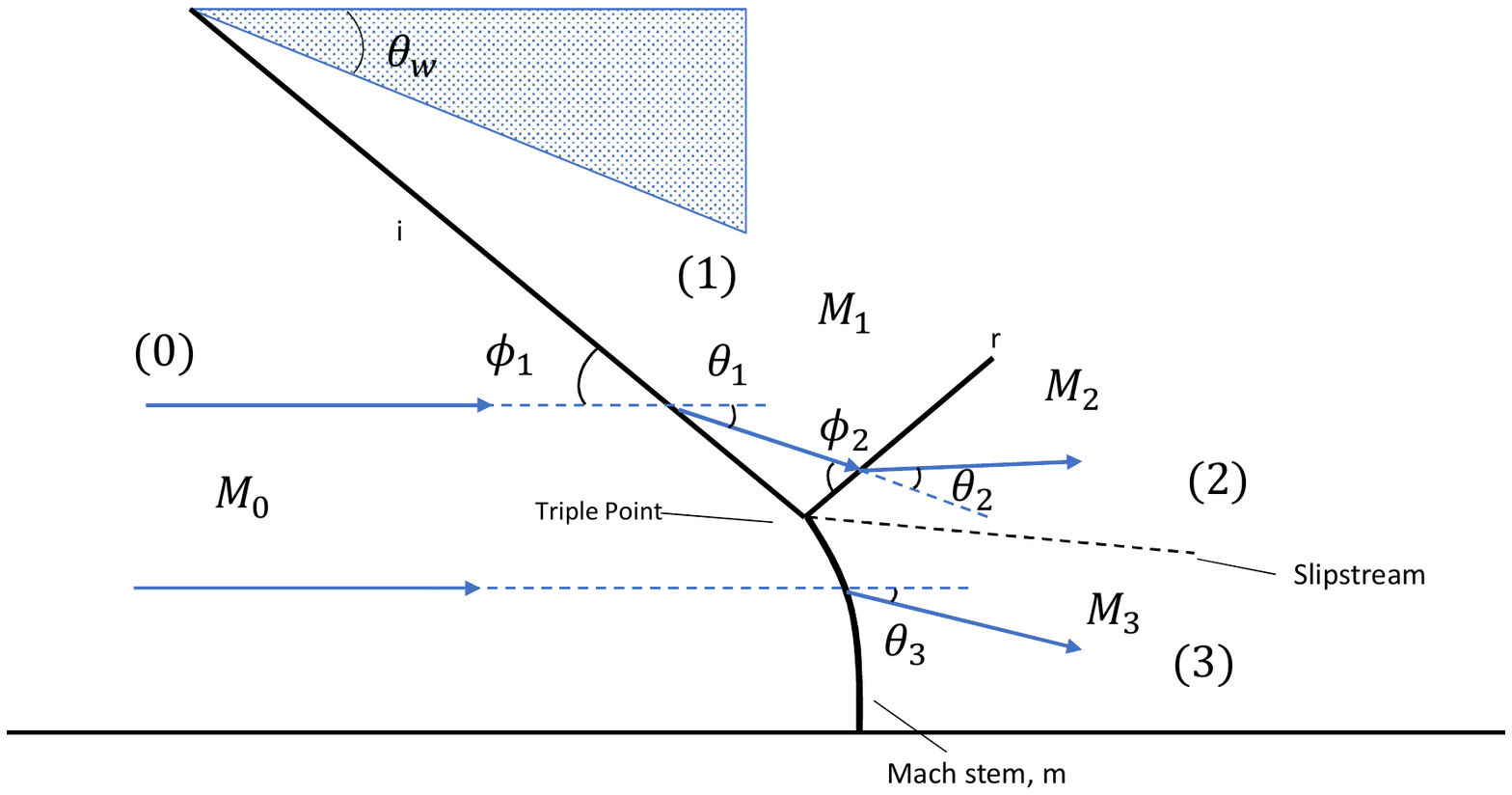}
\caption{\label{fig:mach} Mach reflection configuration for $\theta_w^N(M_0) \leq \theta_w(M_0) < \theta_{max}(M_0)$, where subscript $w$ denotes wedge, $N$ denotes von Neumann angle and $max$ denotes maximal flow turning angle. $M_0$ refers to the in-flow Mach number in the pre-shock flow state.}
\end{figure}

\begin{figure}
\centering
\includegraphics[trim = {0cm, 8cm, 8cm, 5cm}, width = \linewidth]{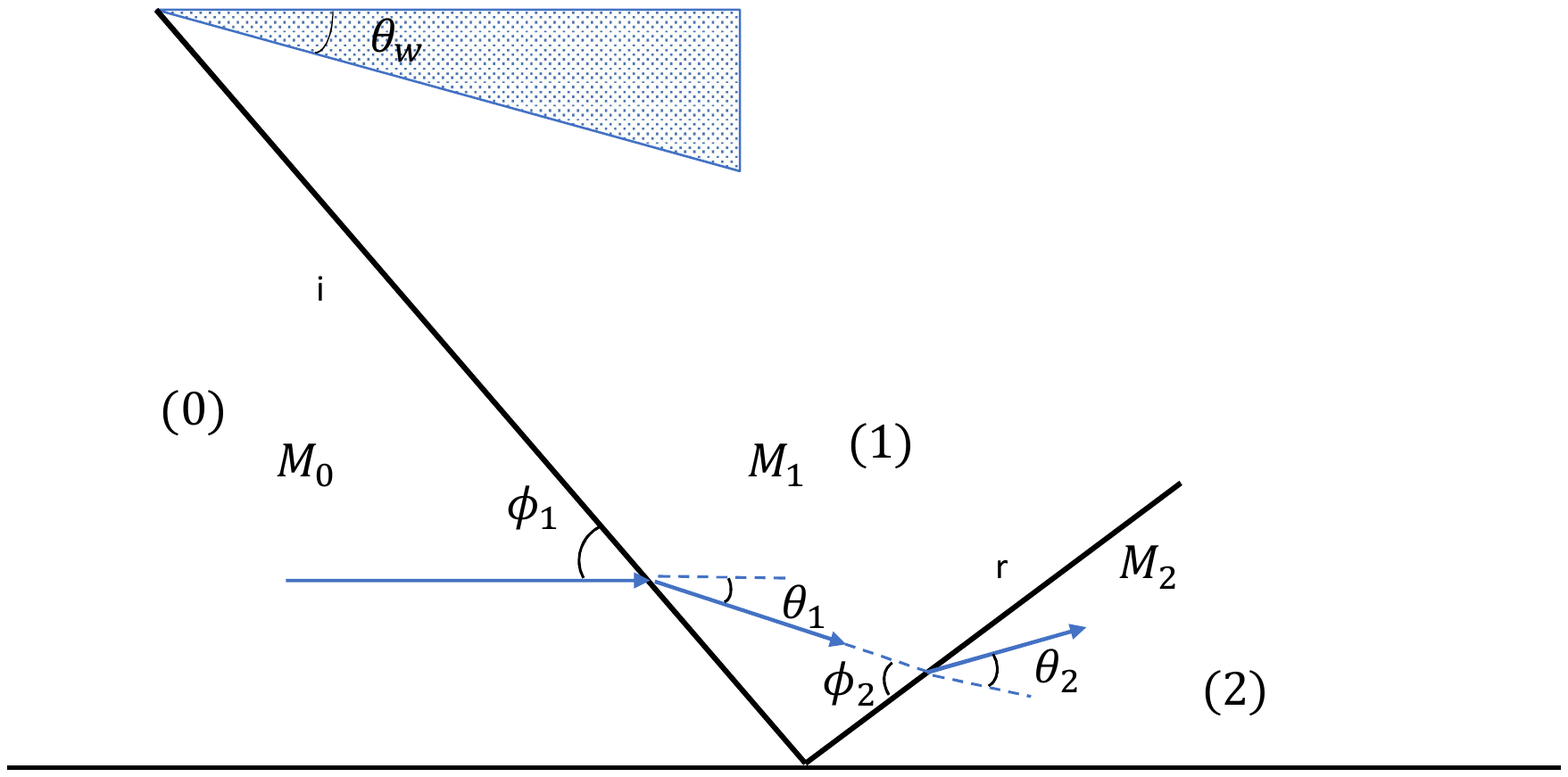}
\caption{\label{fig:reg} Regular reflection for $0 \leq \theta_w(M_0) \leq \theta_{wd}(M_0)$, where subscript $w$ denotes wedge, $N$ denotes von Neumann angle and $d$ denotes detachment angle. $M_0$ refers to the in-flow Mach number in the pre-shock flow state.}
\end{figure}

Shock reflection structures are classified as either a Regular Reflection (RR) or an Irregular Reflection (IR). In steady two-dimensional flows, it is well known that IR takes the form of Mach Reflection (Fig.~\ref{fig:mach}). From the classical two- and three-shock theories \citep{bendor}, the RR occurs when the attached reflected shock can produce a flow deflection angle in the opposite direction that is equal to or greater than that produced by the incident shock (Fig.~\ref{fig:reg}). If the reflected shock is unable to match the deflection angle produced by the incident shock, a three-shock configuration will emerge, called Mach Reflection (MR).
The shock configurations of the RR and IR are fundamentally different, with the Mach stem in the MR typically inciting much higher pressure and temperature downstream than the RR, as well as a subsonic vortex core due to its curvature, particularly in internal flows such as open jets \citep{hornung2000oblique,Hadjadj2009,martelli2020flow}.

It has been well established that there are two main transition criteria for the MR~$\leftrightarrow$~RR transition. These are known as the von Neumann and detachment conditions \citep*{henderson1975experiments,Hornung1979, hornung1982transition,Chpoun1995}. The von Neumann criterion predicts that MR~$\leftrightarrow$~RR transitions occur when the flow deflection angle of the incident shock equals that of the reflected shock, and that their total shock strength must equal to that of a normal shock. If the flow deflection angle (or equivalently, wedge angle $\theta_w $) is lower than the von Neumann angle, MR structures are not possible (see Fig.~16 in \citet{hornung1986regular}); note that this does not include the special case of InMR (Inverted Mach Reflection). The detachment criterion proposes that a MR~$\leftrightarrow$~RR transition occurs when the incident flow deflection angle is greater than the maximum possible turning angle for the reflected shock. For wedge angles greater than the detachment criterion, RR structures are not possible. 
The region between the von Neumann and detachment angle is known as the dual-solution domain, wherein both MR and RR structures are theoretically possible with wedge-angle variation. With the T3 Stalker gun tunnel at the Australian National University, \citet*{Hornung1979} were the first to hypothesise the possibility of a hysteresis process between MR~$\leftrightarrow$~RR transitions in the dual-solution domain. This subsequently led to an innumerable amount of analytical, numerical and experimental work elucidating the mechanisms of the hysteresis process, as detailed by \citet{ben2002hysteresis}. Interestingly, the exact same phenomenon has also been observed within astrophysical jet emissions \citep{yirak2013mach}.

\begin{figure}
\centering
\includegraphics[trim = {5.1cm, 9.7cm, 13cm, 2cm}, width = 0.7\linewidth]{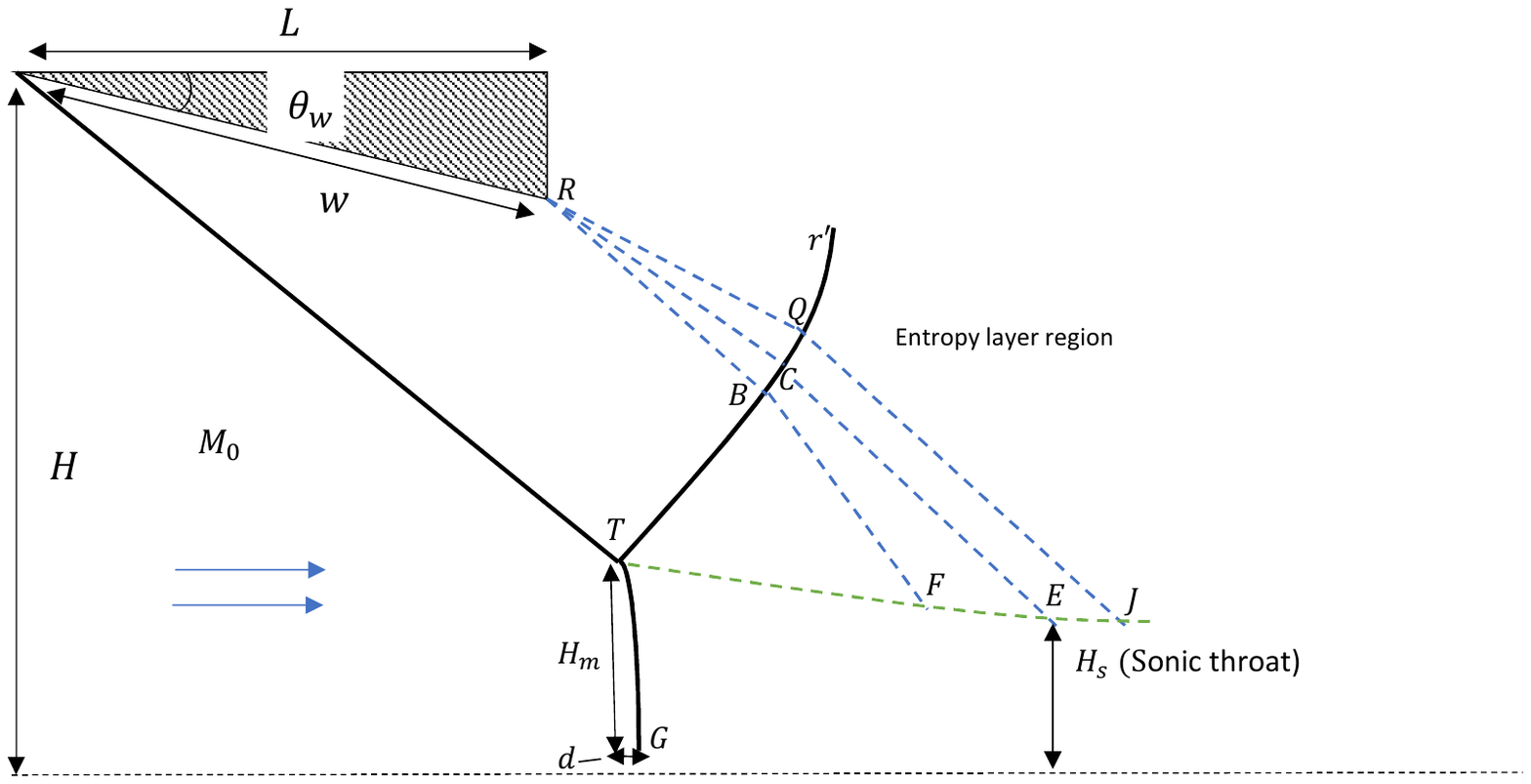}
\caption{\label{fig:bendor} Schematic illustration of the Mach reflection wave configuration and definition of parameters \citep{Li1997}. }
\end{figure}

With regards to the symmetric MR configuration, campaigns to understand its flow topology from a mathematical and analytical standpoint began with \citet{azevedo1989analytic}. The author derived a model that considered the overall flow features of the incident, reflected shock and Mach stem, where the latter is assumed to be straight. Moreover, the slipstream is assumed to be parallel to the reflecting surface up until when the expansion fan reaches it, thereby creating an artificial choking point. \citet{Li1997} provided substantial improvement to the models of \citet{azevedo1989analytic} and \citet{Azevedo1993} (See Fig.~\ref{fig:bendor}). The most important differences between the two are that \citet{Li1997} considered a converging slipstream, and the assumption that the centred expansion characteristic intersects the slipstream exactly at the sonic throat height. They further  accounted for the curvature of each individual component including the shock-expansion interaction involving weak Mach waves at the trailing edge. The latter subsequently generates an entropy layer within which the entropy varies infinitesimally, subsequently curving the reflected shock. Within this configuration, the slipstream curves towards the symmetry plane in the region FE due to the interaction with the right running limiting characteristic CD, precipitating in the artificial choking point at point E, which terminates the subsonic pocket TGFE. Moreover, a second-order curve was further used to model the curvature of the slipstream, as well as the other curved components. 

Following the work of \citet{Li1997}, numerous steady symmetric MR \citep{mouton2007mach,Mouton2008,Gao2010,Bai2017} and asymmetric oMR  (overall Mach Reflection) \citep{Tao2017,Roy2019,Lin2019} analytical models have since been developed, which are similarly capable of predicting the salient organisation of the wedge-induced shocked flowfields. 

In particular, \citet{mouton2007mach,Mouton2008} and \citet{hornung2008some} formulated a symmetric MR model by heuristic analysis of each shock component as straight discontinuities. They adopted \citet{Li1997}'s assumption that the emitted centred expansion waves would interact with the slipstream, resulting in its slight curvature up to the point of the sonic line, wherein it then becomes parallel to the reflecting surface. \citet{Gao2010} and \citet{Bai2017} proposed models where secondary expansion waves were formed along the slipstream, which would eventually transmit upstream to alter the Mach stem height. 

With regards to asymmetric oMR models, the motivation for their derivation is because in most practical engineering applications such as nozzle flows, wedge intake and external flows, reflection of asymmetric shock waves are more likely to occur owing to upstream flow non-uniformities \citep{li1999analytical,ivanov2002,Hadjadj2009,laguarda2021experimental}.  

\citet{Roy2019} suggested a new asymmetric MR model by extending the work in \citet{Li1997}. In their configuration, they included a perturbation angle in the geometrical relations for determining the Mach stem height, which subsequently led to an over-prediction of certain properties when applied to the symmetric MR case. They further devised a method of predicting the subsonic flow properties in the downstream pocket based on averaged flow deflection angles, rather than by using normal shock solutions as in \citet{Li1997}. Thus, the above theoretical considerations allow for extensions to be made to model the Mach stem height in compressible open jets, as detailed in $\S~\ref{sec:2}$.

\section{Shock reflections in over-expanded jets}\label{sec:2}
\begin{figure}
\centering
\includegraphics[trim = 3cm 9.5cm 5cm 1cm,clip,width = \linewidth]{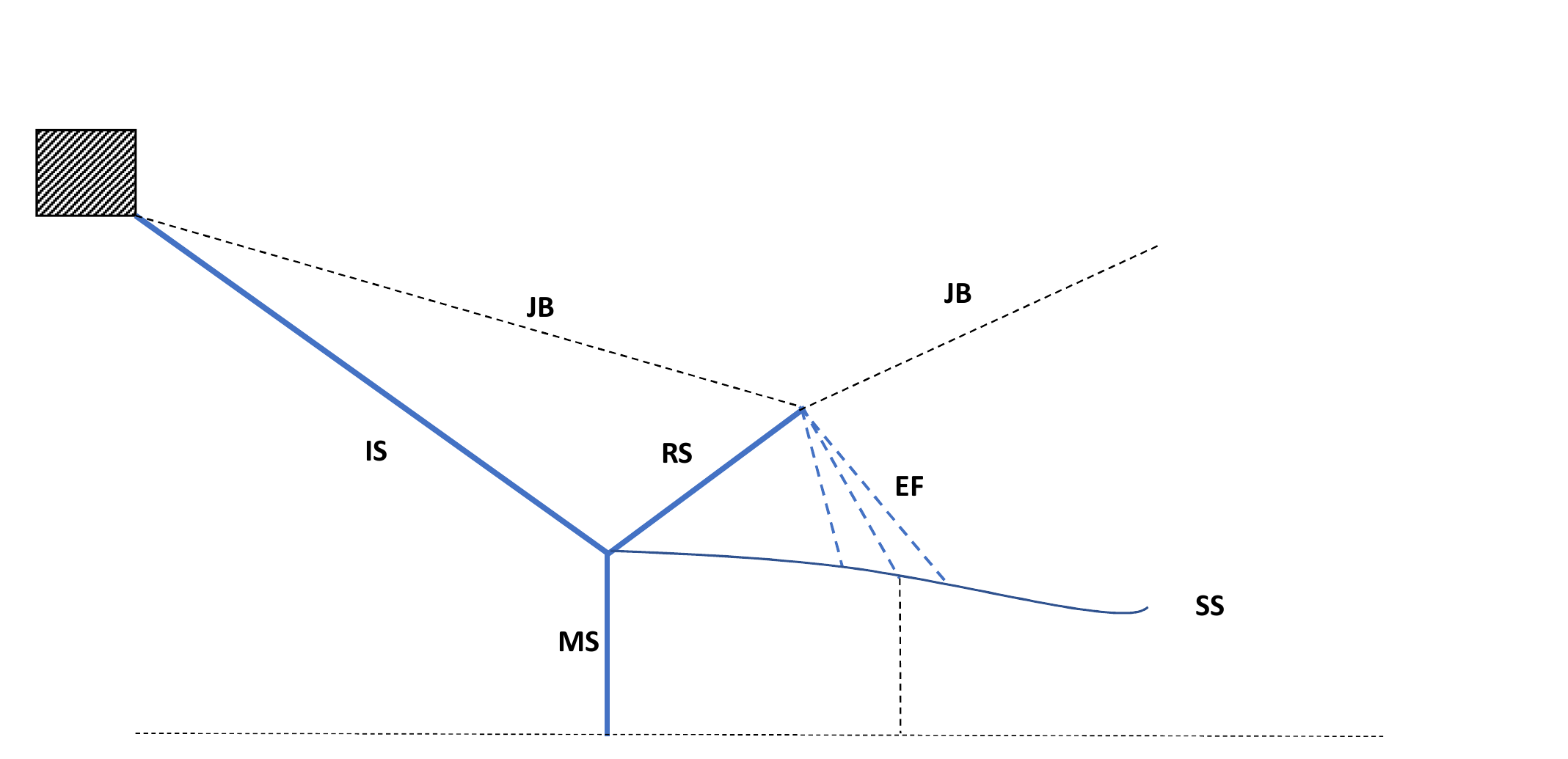}
\caption{\label{fig:jet} General schematic of the Mach Reflection in an over-expanded jet, assuming parallel and uniform flow \citep{frey1999shock,hadjad2004}.}
\end{figure}

Within the context nozzle open jet flows, it has been well documented now that some details of the shock structure within an over-expanded nozzle significantly differ from those of the typical wedge-induced case \citep{hadjad2004, Nasuti2009} ~(see Fig. \ref{fig:jet}). For the flow between two wedges, centred expansion waves emanate from the trailing edge, which subsequently interacts with the reflected shock (RS) to induce a slight curvature \citep{li1996oblique,Li1997}, whereas for over-expanded nozzle flows, a jet boundary (JB) trails above the incident shock (IS), with both JB and IS initially straight. However, when the JB interacts with the reflected shock (RS), the RS is reflected away with via centred expansion waves. This precipitates in multiple rarefactions (Mach waves) that emanate from that point down towards the slipstream. The differing flow features therefore necessitate a slightly different approach in modelling the shock configuration. 


As demonstrated above, many analytical studies have considered the wedge-induced shock reflection for purposes of fundamental understanding of the intrinsic flow configurations, while few investigators have focussed on the theoretical aspects of shock-induced separation within over-expanded planar nozzles. The most prominent attempt at deriving an analytical method to predict parts of the flow topology of the classical MR was conducted by \citet*{chow1975mach}. They approximated the Mach stem height by utilising surface integrals over a strip portion of the subsonic pocket, and further modelled the reflected shock as a second-order curve, and obtained a model for the three-shock structure. However, they did not consider the reflected shock-jet boundary interaction, and its influence on the shape of the slipstream. 

\citet{li1998mach} suggested another model wherein analytical solutions were obtained for the entire shock cell including the expansion wave region, its disturbance and effect on the shape of the slipstream,  as well as the Mach stem height.  \citet{hadjadj1998comparison} applied their model to study the properties of shock curvature within the over-expanded nozzle flow field, wherein they found good agreement with numerical observations. While \citet*{paramanantham2022prediction} recently modified models for the wedge flows \citep{Li1997,mouton2007mach,Bai2017} for parametric analyses of Mach stem height and growth rates in over-expanded jets. To our knowledge, no other theoretical frameworks have been developed to predict the morphology of shock configurations occurring within an over-expanded jet. 

\citet*{hadjad2004} and \citet*{shimshi2009viscous} numerically studied the near-flow field of a shock-containing jet by use of viscous approaches with Reynolds-averaged Navier Stokes (RANS) turbulence models and identified a hysteresis pattern within the dual-solution domain (cf. Fig. 16 in \citet{hornung1986regular}), in exact agreement with the von Neumann and detachment criteria for wedge-induced shock reflections. While other investigators have predominantly concentrated on the side-load behaviour due to shock-generated transient wall-pressure fluctuations, aero-acoustic noise generation and shock-driven turbulence \citep{baars2012wall,jaunet2017wall,martelli2020flow,bakulu2021jet}. Moreover, a primary challenge associated with experimental investigations on the shock structure is the lack of optical access within the nozzle especially when the flow is asymmetric, which complicates the application of flow visualisation techniques. Therefore, this warrants the development of analytical models in combination with numerical methods to predict the internal flow structure interactions.

\begin{figure}
\centering
\includegraphics[trim = {5cm, 7cm, 13cm, 2.2cm}, width = 10cm]{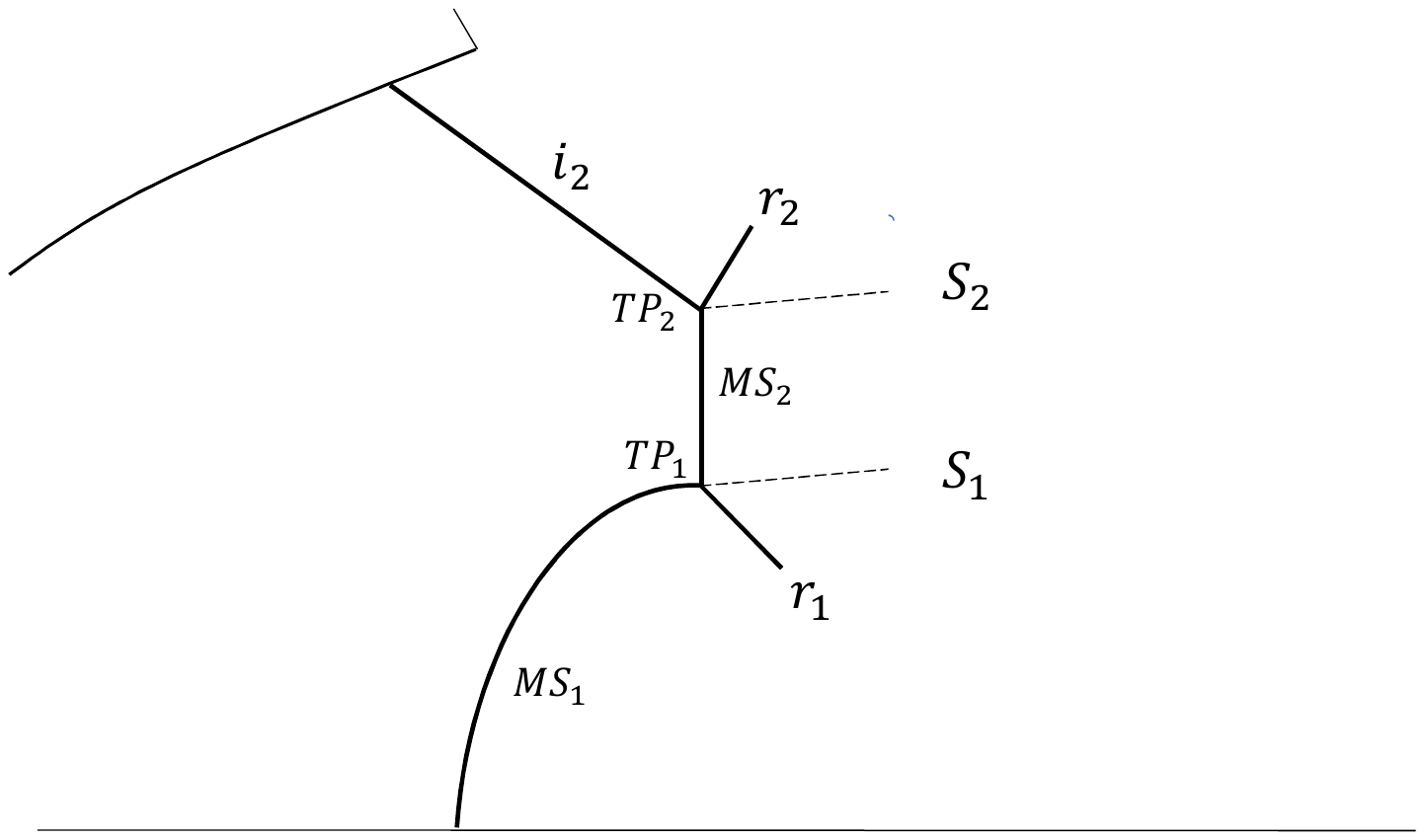}
\caption{\label{fig:cap}Schematic illustration of the peculiar cap-shock Mach reflection in over-expanded supersonic jets, the curvature of the central Mach stem ($MS_1$) is due to upstream flow non-uniformities, produced without the existence of an internal shock, it is the so-called Type-II non-uniform Mach reflection \citep{Nasuti2009,Hadjadj2009}. }
\end{figure}

Within supersonic jets, one rather unique type of shock structure that occurs during shock-induced separation is the cap-shock pattern with no internal shock arising from steep gradients near the nozzle throat, or the so-called Type-II cap shock interference \citep{Nasuti2009,Hadjadj2009} (Fig.~\ref{fig:cap}), whose classification is based on \citet{edney1968anomalous,edney1968effects}. The salient features are markedly different from all other observed configurations such as the classical MR, oMR or RR. The cap-shock phenomenon, or presently called cap-shock Mach reflection (csMR), as it resembles the classical MR, is documented in detail by \citet{Hadjadj2009} and \citet{Nasuti2009}, wherein the authors attempted to provide an explanation of its origins. To date, there seems to be no analytical study on this configuration and present models can not be readily applied to these structures.

Based on the current understanding of shock reflection phenomena in supersonic jets, the observed structure, albeit uncommon, occurs without the presence of an internal shock and only where there is a highly non-uniform upstream flow gradient. Thus, it may appear within contoured wall nozzles such as the TIC (Truncated Ideal Contour) or planar minimum length nozzles (MLN), which are all generally referred as MLNs~\citep{nasuti2005influence,Nasuti2009,Hadjadj2009}, which are designed based on the method of characteristics (MoC), and truncated to only include the straightening section of the contour, which can result in a non-uniform inflow.   

In consideration of all this, the present study has been conducted to investigate the characteristics of the csMR from an analytical and numerical viewpoint. The remainder of this work is organised as follows: first, a novel analytical framework for the prediction and parametric analysis of the asymmetric cap-shock Mach reflection within an over-expanded nozzle is formulated, which includes consideration of the pre- and post-shock flow properties near the triple point juncture, the shape and morphology of the Mach stem with respect to the subsonic pocket, and the expansion-slipstream interaction. Secondly, a Riemann computational fluid dynamics (CFD) shock-capturing numerical scheme is employed for purposes of comparisons with the above analytical model, using a self-made method of characteristics code for the supersonic minimum length nozzle design.

\section{Analytical Framework}\label{sec:types_paper}

This section contains detailed information about the formulation of the present non-uniform Type II cap-shock Mach reflection model. The entire algorithm was coded with the use of Python, particularly with the SciPy module, which utilises the modified Powell hybrid Jacobian method to find numerical solutions of nonlinear systems of equations \citep{more1980user}. 


\subsection{Base Framework for the Type-II non-uniform Cap-Shock Mach Reflection}
The main body of the analytical framework extends upon the symmetric and asymmetric MR models of \citet{Li1997}, \citet{li1998mach} and \citet{Roy2019}, where \citet{Li1997} and \citet{Roy2019} are for wedge-induced oblique shock reflection, while \citet{li1998mach} is for the symmetric MR in an over-expanded jet. They have been derived by improving upon the earlier methods of \citet{azevedo1989analytic} and \citet{Azevedo1993}. The extensions made here mainly lie in the development of a closed, coupled system of equations that solve for the asymmetric Mach stem height in the upper and lower domains of an open jet. This is crucially made possible by incorporating a non-uniform upstream Mach number behind the triple point juncture, as detailed in the later sections.

\begin{figure}
\centering
\includegraphics[trim = 0cm 2cm 6cm 2cm ,width = \linewidth, ]{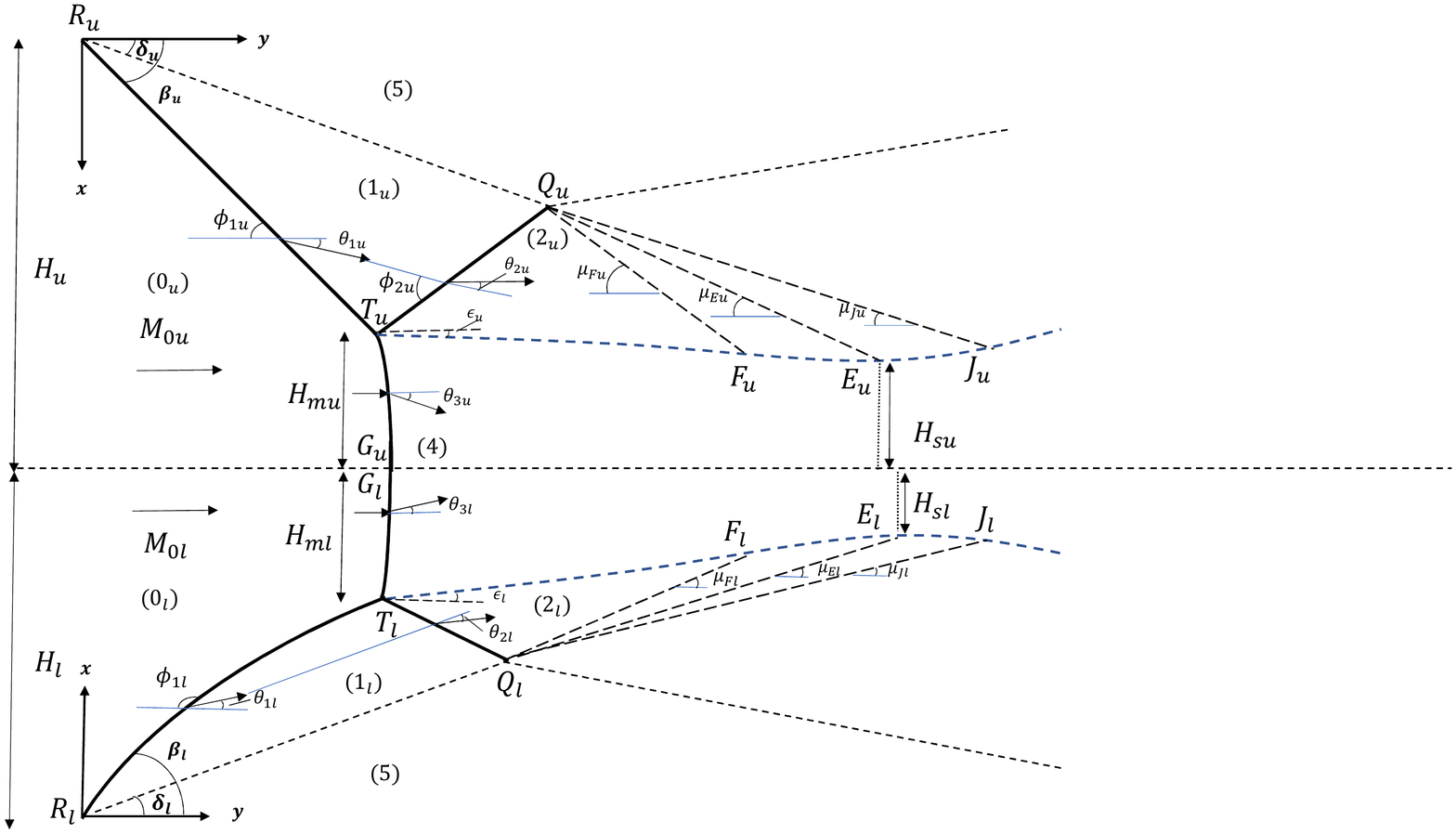}
\caption{\label{fig:crazy} Schematic illustration of the Type-II asymmetric cap shock Mach reflection with the relevant parameters.}
\end{figure}

A schematic illustration of the present model is given in Fig.~\ref{fig:crazy}. Note that the present model is for a DiMR--DiMR configuration, wherein both the left- and right-running shock waves deflect streamlines towards the $x$-axis. The Mach stem is therefore a right-running shock wave, rather than a left-running one in the case of \citet{Roy2019}, where $\theta_{3u}$ would otherwise be negative to adhere with the sign conventions. The upper domain corresponds to the symmetric MR model proposed by \citet{li1998mach}, with the inclusion of an extra point $J$ here not considered in their configuration. Similar to the wedge-induced symmetric MR model of \citet{Li1997}, \citet{li1998mach}'s model systematically solves for all unknown parameters in the governing shock relations by relating pre- and post-shock flow variables. Thus, upstream flow properties such as the Mach number $M_0$, pressure $P_0$, density $\rho_0$, and acoustic speed $a_0$, allow all geometrical features to be computed via nonlinear sets of coupled equations. The geometrical features include the interaction of the jet boundary (JB) characteristic and the reflected shock wave as in point Q in Fig.~\ref{fig:crazy}, which results in a Prandtl-Meyer wave region QFEJ, the curvature of the slipstream TFEJ, and the flow properties in the subsonic pocket downstream of the Mach stem TG. The curvature of the Mach stem is then also modelled as a second-order curve with boundary conditions placed at its ends \citep{Li1997,Roy2019}. Moreover, the point~E is assumed to be parallel to the symmetry plane, where flow re-choking occurs, thereby terminating the subsonic pocket TEG. 

In this model, we do not impose a universal coordinate system. Instead, all flow properties are computed for two different sets of coordinates as denoted in Fig.~\ref{fig:crazy}, but continuity at the intersection of the Mach stems is used to close the governing equations. The introduction of asymmetry due to upstream flow non-uniformities is achieved through the use of differing upstream properties in the upper and lower regions, as detailed further in Sec. \ref{sec:triple}.

Additionally, it must be noted that \citet{Gao2010}, \citet{Bai2017} and \citet{Lin2019} considered secondary wave effects across the slipstream, where it influences the position of the choking point with respect to the subsonic pocket, subsequently affecting the Mach stem height. However, if the interaction between secondary expansion waves with the oblique shocks aft of the first shock cell are ignored, with the understanding that the impedance differences result in negligible reflection intensity \citep{Li1997,henderson1989refraction}, \citet{Li1997} showed that purely geometrical considerations need to be utilised to obtain a parametric description of the entire shock structure including the Mach stem height $H_m$.

\subsubsection{Triple Point Juncture}\label{sec:triple}
We first consider the vicinity near the triple point juncture. The classic von Neumann three-shock theory \citep{von1943oblique,von1945refraction,bendor} is applied to each triple point, which bifurcates into a confluence of four types of Rankine-Hugoniot (R-H) discontinuities, the incident shock, reflected shock, Mach stem, and the tangential discontinuity known as a slip stream. 
The following equations relate the flow properties behind an oblique shock, to the ones ahead of it. These variables are the Mach number $M$, pressure $P$, flow deflection angle $\theta$, shock angle $\phi$, density $\rho$ and acoustic speed $a$. More details of 3-ST can be found in \citet{Li1997} and \citet{bendor}, which are derived based on the Rankine-Hugoniot conservation equations.

In typical three-shock theory (3-ST) notation~\citep{bendor}, these relations are,
\begin{equation}\label{eqn:2.10}
\left.\begin{array}{rl}
M_{j} & =F\left(M_{i}, \phi_{j}\right) \\
\theta_{j} & =G\left(M_{i}, \phi_{j}\right) \\
P_{j} & =P_{i} H\left(M_{i}, \phi_{j}\right) \\
\rho_{j} & =\rho_{i} E\left(M_{i}, \phi_{j}\right) \\
a_{j} & =a_{i} A\left(M_{i}, \phi_{j}\right)
\end{array}\right\} \equiv \xi_{i} \Pi\left(M_{i}, \phi_{j}\right),
\end{equation}
\begin{equation}
    F(M, \phi)=\left[\frac{1+(\gamma-1) M^{2} \sin ^{2} \phi+\left[\frac{(\gamma+1)^{2}}{4}-\gamma \sin ^{2} \phi\right] M^{4} \sin ^{2} \phi}{\left[\gamma M^{2} \sin ^{2} \phi-\frac{\gamma-1}{2}\right]\left[\frac{\gamma-1}{2} M^{2} \sin ^{2} \phi+1\right]}\right]^{1 / 2},
\end{equation}
\begin{equation}
G(M, \phi)=\arctan \left[2 \cot \phi \frac{M^{2} \sin ^{2} \phi-1}{M^{2}(\gamma+\cos 2 \phi)+2}\right],
\end{equation}
\begin{equation}
H(M, \phi)=\frac{2}{\gamma+1}\left[\gamma M^{2} \sin ^{2} \phi-\frac{\gamma-1}{2}\right],
\end{equation}
\begin{equation}
    E(M, \phi)=\frac{(\gamma+1) M^{2} \sin ^{2} \phi}{(\gamma-1) M^{2} \sin ^{2} \phi+2},
\end{equation}
\begin{equation}\label{eq2.15}
A(M, \phi)=\frac{\left[(\gamma-1) M^{2} \sin ^{2} \phi+2\right]^{1 / 2}\left[2 \gamma M^{2} \sin ^{2} \phi-(\gamma-1)\right]^{1 / 2}}{(\gamma+1) M \sin \phi}.
\end{equation}

The pre- and post-shock media depend only on the upstream Mach number and shock angle. Thus, across the first incident shock (see Fig.~\ref{fig:crazy}) ($R_u T_u$),
\begin{equation}
\xi_{1_u} = \xi_{0_u} \Pi(M_{0_u}, \phi_{1_u}),
\end{equation}
across the first reflected shock ($T_u Q_u$),
\begin{equation}
\xi_{2_u} = \xi_{1_u} \Pi(M_{1_u}, \phi_{2_u}),
\end{equation}
across the upper half segment of the convex Mach stem ($T_u G_u$),
\begin{equation}
\xi_{3} = \xi_{0} \Pi(M_{0_u}, \phi_{3_u}),
\end{equation}
across the lower central convex Mach stem ($T_l G_l$),
\begin{equation}
\xi_{1_l} = \xi_{0_l} \Pi(M_{0_l}, \phi_{1_l}),
\end{equation}
across the second reflected shock ($T_l Q_l$),
\begin{equation}
\xi_{2_l} = \xi_{1} \Pi(M_{1_u}, \phi_{2_l}),
\end{equation}
and across the lower half segment of the Mach stem ($R_l T_l$),
\begin{equation}
\xi_{3_l} = \xi_{0_l} \Pi(M_{0_l}, \phi_{3_l}).
\end{equation}
We close the governing equations with standard shock theory solutions and with the requirement of pressure continuity along the slipstreams, given by
\begin{equation}
\theta_1= \delta \:, \: P_3 = P_2 \:, P_1= P_{\infty} = P_5 \:, \: \theta_3 = \theta_1-\theta_2 \:, \: \beta = \phi_1 \:, \: \theta_3 = \epsilon,
\end{equation}
where the subscripts $u$ (upper domain) and $l$ (lower domain) are implied. Thus, considering both the upper and lower domains (i.e., splitting the Mach stem $MS_2$ into two parts), classical shock theory yields 36~equations that are solved for 36~unknowns. Just considering the upper domain, the 18~unknowns are $M_i$, $P_i$, $\phi_i$, $\theta_i$, $\rho_i$, and $a_i$, with $i=1,2,3$. Note that in the case of over-expanded flow, $\theta_1$ is not known \textit{a priori}, since no wedge angle exists. Instead, the value $P_{\infty} = P_5 = P_1$, where $P_{\infty}$ is the ambient back pressure, gives the necessary condition to complete the governing set of equations \citep{li1998mach}.

Note that this method is distinct from other asymmetric oMR models in the sense that $M_0$ is non-uniform, and so differs in the upper and lower domains by assumption. This is due to the simple fact that asymmetric shock reflection occurring in nozzle flowfields are due to upstream flow non-uniformities, rather than difference in wedge angles as in the case of wedge-induced shock reflection. The non-uniformity in Mach number also represents a more realistic situation of how asymmetric shocks are produced in reality. Moreover, approximating the non-uniformity of Mach number as two discrete values is also valid under the assumption that the shock structure is only slightly asymmetric, which is the case in most circumstances.

Other than that, provided $M_0$ in both regions are known, the remaining flow parameters; namely, $P_0$, $\rho_0$, $a_0$ in each domain are computed by quasi-one dimensional relations, which are related to the Mach number $M_0$ and $\gamma$ in their respective regions.

\subsection{Mach stem curvature and shape}
\citet{Li1997} showed that if we assume the Mach stem is only slightly curved, it is possible to model its shape as a second-order curve, defined implicitly with boundary conditions placed at its ends (slope and coordinates). This assumption of slight curvature is valid for symmetric MR and asymmetric oMR \citep{Roy2019}. However, in the case of the csMR configuration, it is clear that the central Mach stem is significantly curved and cannot readily be modelled using the method described. On the other hand, \citet*{Tan2006} proved, using small-disturbance theory, that the Mach stem can be better approximated as a circular arc for a non-uniform flow field. For the central Mach stem, we combine these two concepts to derive a circular arc approximation by considering the circle in the osculating plane that traces the midpoint of \citet{Li1997}'s second-order curve. Details of the derivation, along with the final expression are provided in Appendix~A, where the resulting form results in a considerably simpler expression than that of \citet{Li1997}, which allows ease of use.

Following \citet{Li1997,li1998mach}, the shape of the other slightly curved Mach stem is obtained via a second-order curve as
\begin{equation}
    y = -1/2 \cot\phi_4(x-H)^2/H_m + (H-H_m)\cot\beta + 1/2 H_m \cot\phi_4.
\end{equation}
$\phi_4$ here is defined in terms of averaged deflection angles based on \citet{Roy2019} and \citet{Lin2019}, it is introduced to compute the flow properties through the asymmetric subsonic pocket as discussed in Sec. \ref{sec:2.4}. Further details of the second-order curve approximation is provided in \citet{Li1997}, where the result is obtained based on the geometrical configuration detailed in Appendix~B. Note that this second order curve assumes a slope at point $G$ being orthogonal to the freestream flow. This is valid for DiMR-DiMR structures but would not be applicable to InMR structures since additional constraints must be imposed.

\subsection{Flow properties through the subsonic pocket (region TEG)} \label{sec:2.4}
The flow within the jet domain demarcated by the symmetry plane and the slipstream is subsonic. Theoretically, the flow in this configuration is impossible to model exactly, as its properties can be determined by downstream conditions \citep{Chpoun1999,ben1999influence}. 

However, it is well understood that approximate expressions can be obtained by considering the flowfield only near the triple-point juncture by utilising mean deflection quantities \citep{Tao2017,Roy2019,Lin2019}. Based on our results obtained from numerical simulations of the non-uniform cap shock pattern, we conclude that flow deflections ahead of the Mach stem require consideration of average deflection angles. Similar to \citet{Tao2017}, we introduce an averaged deflection parameter at the Mach stem as,
\begin{equation}
    \theta_{4_{avg}} = \theta_{avg} = \frac{\theta_{3u} + \theta_{3l}}{2},
\end{equation}
where the positive signs are in-keeping with the convention adopted by \citet{Roy2019}, whose configuration was for a DiMR-InMR overall Mach reflection structure. Following
\citet{Li1997} and \citet{Roy2019}, we henceforth assume the flow proceeds quasi one-dimensionally, until the sonic line at the throat, which yields the well-known isentropic ``Area-Mach" relation given by
\begin{equation}\label{eqn2.24}
  \frac{H_{m}}{H_{s}}=\frac{1}{\bar{M}}\left[\frac{2}{\gamma+1}\left(1+\frac{\gamma-1}{2} \bar{M}^{2}\right)\right]^{\gamma+1 /(2(\gamma-1))},
\end{equation}
with
\begin{equation}
\bar{M}=\bar{u} / \bar{a}.
\end{equation} 
Under a first-order approximation, it follows that
\begin{equation}
\bar{u}=\frac{1}{H_{m} \bar{\rho}} \int_{0}^{H_{m}} \rho \boldsymbol{u} \cdot \boldsymbol{e}_{x} \mathrm{~d} y=\frac{1}{2 \bar{\rho}}\left(\rho_{3} u_{3} \cos \theta_{3}+\rho_{4} u_{4} \cos\theta_4\right),
\end{equation}
\begin{equation}
\bar{a}=\frac{1}{2}\left(a_{3}+a_{4}\right), \quad \bar{\rho}=\frac{1}{2}\left(\rho_{3}+\rho_{4}\right),
\end{equation}
\begin{equation}
\bar{M}=\frac{2\left(\rho_{3} u_{3} \cos \theta_{3}+\rho_{4} u_{4}\cos \theta_4\right)}{\left(\rho_{3}+\rho_{4}\right)\left(a_{3}+a_{4}\right)},
\end{equation}
where $u_3 = M_3a_3$ and $u_4 = M_4a_4$. The quantities $M_3$, $\rho_3$ and $\theta_3$ are obtained from 3-ST, and $M_4$, $\rho_4$, and $a_4$ are computed from oblique-shock theory with the strong-shock solution assumption ($M_4 < 1$),
\begin{equation}
 \xi_{4} = \xi_{4} \Pi(\bar{M}_{0}, \phi_4).
\end{equation}
where $\bar{M}_0 = (M_{0u} + M_{0l})/2$. This yields an additional 5~equations with 5~unknowns, which allows all the averaged flow quantities downstream of the Mach stem to be recovered. Thus, different from the wedge flow case, the non-uniformity in the Mach number $M_0$ implies that an averaged upstream Mach number should be considered.

Note that Eq.~(\ref{eqn2.24}) has two unknowns, namely $H_m$ and $H_s$. Therefore, in order to close the governing set of equations, a relation for $H_m$ is required, which is given through a set of geometrical relations outlined in Appendix~B. The quasi-one-dimensional assumption and the geometrical relations result in a total of 33~equations that are solved for 33~unknowns, necessarily determining the locations of all points on Fig.~\ref{fig:crazy}. Further details on the calculations are provided in Appendix~B. 

\subsection{Interaction of the jet boundary (JB) and the reflected shock (RS)}
As mentioned earlier, the interaction between the reflected shock and centred expansion waves for an over-expanded nozzle differs slightly from the wedge-induced case in the sense that for wedge flows, expansion fans emerge from the trailing edge, thereby inducing curvature of the reflected shock. However, for nozzles, no such phenomenon occurs. Instead, the right-running JB characteristic intersects the RS, which subsequently generates reflected expansion waves that interact with the slipstream. This interaction is modelled as a simple wave process as in \citet{li1998mach}.

Thus, in the present case, the region TQFE is modelled with the Prandtl-Meyer relation, with net flow deflection angle $\epsilon$, given by,
\begin{equation}
\epsilon=v\left(M_{C}\right)-v\left(M_{2}\right),
\end{equation}
\begin{equation}
    v(M)=\left(\frac{\gamma+1}{\gamma-1}\right)^{1 / 2} \arctan \left[\frac{(\gamma-1)\left(M^{2}-1\right)}{\gamma+1}\right]^{1 / 2}-\arctan \left(M^{2}-1\right)^{1 / 2},
\end{equation}
\begin{equation}
\chi\left(M_{i}, M_{j}\right)=\left[\frac{2+(\gamma-1) M_{i}^{2}}{2+(\gamma-1) M_{j}^{2}}\right]^{\gamma /(\gamma-1)},
\end{equation}
with the pressure $P_C$ computed as
\begin{equation}
    P_C = P_2\chi(M_2,M_E).
\end{equation}
The subscript $u$ and $l$ are similarly implied as in the preceding discussions.
We consider region QEJ in a similar fashion, i.e., the P-M relation is applied again with the net flow deflection angle $\alpha$, thus yielding
\begin{equation}
\alpha = v\left(M_{J}\right)-v\left(M_{E}\right),
\end{equation}
\begin{equation}
\quad P_{J} = P_{E} \chi\left(M_{E}, M_{J}\right),
\end{equation}
where $\epsilon$ and $M_2$ are known from earlier, so this set constitutes 8~equations with 10~unknowns (both upper and lower domains), namely $\alpha$, $M_J$, $M_E$, $P_C$ and $P_J$, which is incomplete. Thus, we close the set by imposing the appropriate boundary condition $\alpha = \theta_2$, allowing the remaining variables to be computed.



\section{Numerical Simulations} \label{sec:simulations}

Here we provide details on the methods behind the numerical simulations, which will enable direct comparisons with the predictions of the analytic model.

\subsection{Governing Equations}
All simulations carried out in this study have been conducted with the general purpose CFD code called Fluent, which has previously been well-validated for shock-capturing purposes \citep{shimshi2009viscous,filippi2017supersonic,surujhlal2019three,sugawara2020three,janardhanraj2021insights}. The Navier-Stokes equations for 2D compressible flow of calorically perfect single-phase gas are employed to model the open jet flow.

Within shock-capturing schemes, there is a known false diffusion problem called the ``carbuncle phenomena'', wherein anomalous solutions can result that are not entropy-consistent \citep{quirk1997contribution,pandolfi2001numerical}. To avoid this, a Godunov-type approximate Riemann solver is employed with finite volume-conservative discretisation of the governing Euler (inviscid) equations, on an unstructured quadrilateral grid. 

The convective and diffusive fluxes are spatially discretised using the total-variation-diminishing (TVD) second-order upwind scheme \citep{quirk1997contribution}. Numerical fluxes are calculated with the AUSM+ Flux Vector Splitting (FVS) method of \citet{liou1996sequel}, with the Green-Gauss interpolation of cell-centred values to face values for piece-wise linear reconstruction of the preconditioned system \citep{mavriplis2003revisiting}. The \citet{barth1989design} slope limiter is imposed to prevent spurious oscillations, where $\Phi_f$ denotes the general flow variable, which is obtained through the cell-face interpolation scheme, and the gradient is limited by the Minmod (Minimum modulus) function \citep{zhang2015review}. 

To advance the solution in time, we use the implicit-Euler, second-order, symplectic time-integrator to write the governing equations in ``delta form". A Lower-Upper Symmetric Gauss-Seidel (LU-SGS) algorithm \citep{jameson1981implicit} is called to obtain $\Delta \mathbf{Q_i}$ for each cell $i$ and conservative (primitive) variable $\mathbf{Q}$, yielding $\mathbf{Q}_{i}^{n+1}=\mathbf{Q}_{i}^{n}+\Delta \mathbf{Q}_{i}$ for each iteration from time step $n$ to $n+1$. Additionally, local time-stepping is employed to accelerate steady-state convergence.

This adaptive time step is computed from the Courant-Friedrichs-Lewy (CFL) condition,
\begin{equation}
\Delta t=\mathrm{CFL}\frac{2 V}{\sum_{f} \lambda_{\mathrm{f}}{ }^{\max } A_{f}},
\end{equation}
where $\lambda_{\mathrm{f}}{ }^{\max }$ denotes the maximal eigenvalues of local face values in the preconditioned conservative linear system, V and A are the local cell volume and face area respectively. $\mathrm{CFL} \leq 1$ is a safety factor to guarantee stability of the numerical solution in time. Here the CFL number stabilised at near unity for all computations.

AMR (Adaptive Mesh Refinement) techniques \citep{Berger1989} based on a dynamic density gradient criterion are also applied to ensure the shocks are resolved with at least 10~computational cells (longitudinally) within them. The code is parallelised using the MPI library and run via a batch script distributing over Gadi's Intel Xeon Scalable Cascade Lake consisting of 240~compute cores (5~nodes with 48~cores and 192~GB memory per node) at the National Computational Infrastructure (NCI).

Depending on the grid density and number of AMR levels, inviscid shock-capturing simulations can be rather computationally intensive, because of the lack of explicit viscosity to diffuse large gradients over the grid stencil arrangement. In the present study, the computation took about 30~CPU core hours for residuals to flatten.

\subsection{Flow Geometry and Computational Domain}
The geometry is a conventional planar 2D Mach~3 minimum length convergent-divergent nozzle. In order to ensure a uniform free-stream Mach number at the exit region of the nozzle, the contour in the divergent section was constructed via the use of a self-made method of characteristics (MoC) python script, based on the design in \citet{emanuel1986gasdynamics}. With the use of the MoC code, the nozzle geometry is completely prescribed. The height of the throat is $h = 2 \times 10^{-3}\,$m. The nozzle exit-throat area ratio (AR) ($A_e/A^* = 4.2$) results in the design Mach number $M = 3.0$. The nozzle length extends $12h$ in the longitudinal direction and $4.2h$ in the transverse direction. The computational domain extends far behind the nozzle to about $L = 50h$, where a pressure outlet boundary condition is employed in the far-field. The inlet pressure is kept constant at $P_i = 290\,$kPa, and the nozzle pressure ratio (NPR) is varied by changing the outlet back pressure in small increments. Moreover, because of the symmetric nature of the flow field, a symmetry plane boundary condition is applied at the axis.


\subsection{Solution Procedure}
Preliminary numerical computations reveal that although the MLN has been designed with a particular uniform design Mach number, this value is never achieved; resulting in a highly non-uniform flow profile from the symmetry plane to the nozzle lip. At a certain narrow $NPR$ range, the non-uniformity becomes so pronounced that the shock structure undergoes an abrupt transition from the typical Free Shock Separation (FSS) structure \citep{Chen1994,Hadjadj2009,Hadjadj2015,martelli2020flow} to the Type-II non-uniform csMR. Therefore, owing to the MLN design and upstream flow non-uniformities, the planar jet can also yield these peculiar structures. This was similarly observed in numerical simulations by \citet{gross2004numerical}, \citet{gross2004numericalhot}, \citet*{nasuti2005influence}, and in experiments by \citet{reijasse2001flow} for axisymmetric jets, where the structure oscillates to and fro with minor variations of the nozzle pressure ratio (NPR). In our case, once the Type-II csMR pattern is adequately captured within the narrow $NPR$ range, we perform a grid resolution study as detailed in Sec.~\ref{sec:results}

For shock reflection visualisations, the \citet{lovely1999shock} shock detection parameter is used,
\begin{equation} \label{eqn:lovely}
    M_n = \frac{\textbf{V}\cdot\nabla p}{a|\nabla p|},
\end{equation}
where $M_n$ denotes the normal Mach number. 

\section{Results and Discussion}\label{sec:results}
\begin{figure}

    \centering
\includegraphics[trim = {1cm, 1.5cm, 2cm, 1cm}, width = \linewidth]{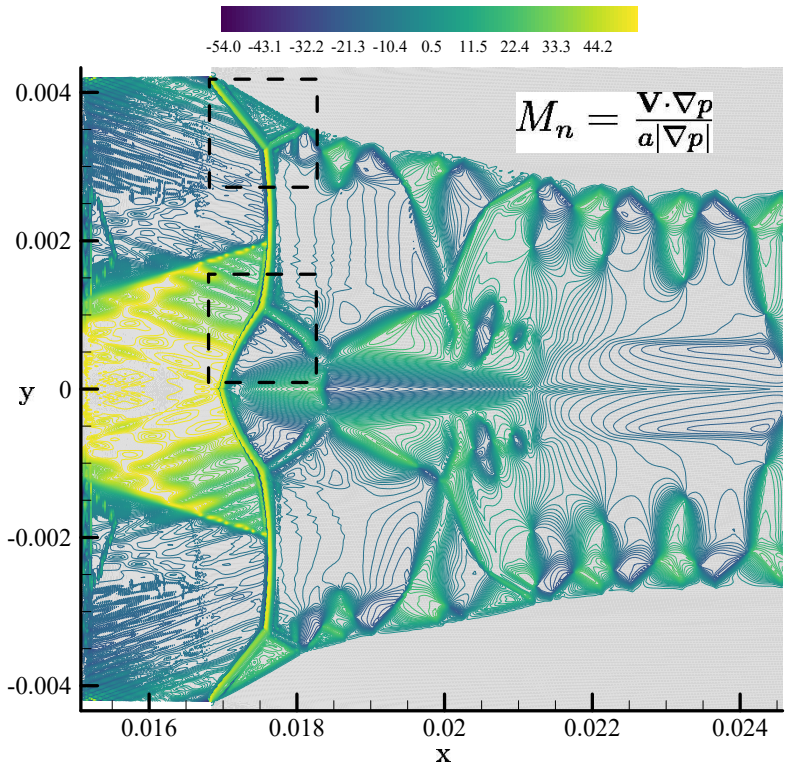}
  \caption{Iso-contour of \citet{lovely1999shock}  shock-detection parameter (Eqn. \ref{eqn:lovely}) for the Type~II csMR configuration outside the over-expanded planar jet (NPR $\approx$ 6.04) with 8~levels of AMR, yielding a grid density of about $ 1235 \times 1100$ cells. Square boxes highlight the features of the first shock cell, which is in full qualitative agreement with the peculiar configuration detailed in \citet{Nasuti2009} and \citet{Hadjadj2009}. Note the fact that the incident shock from the nozzle lip is straight instead of curved, since the flow is planar rather than axisymmetric. Beyond the first shock cell, the salient features include the so-called shock train or {\it supersonic tongue} \citep{hagemann,babinsky2011shock}, where a rapid succession of shock and expansion waves are formed due to the repeated steepening and divergence of Mach lines. }
\label{fig:nice}
\end{figure}
\begin{figure} 
    \begin{subfigure}{0.31\textwidth}
        \centering
        \includegraphics[height = 6cm, width =  \linewidth]{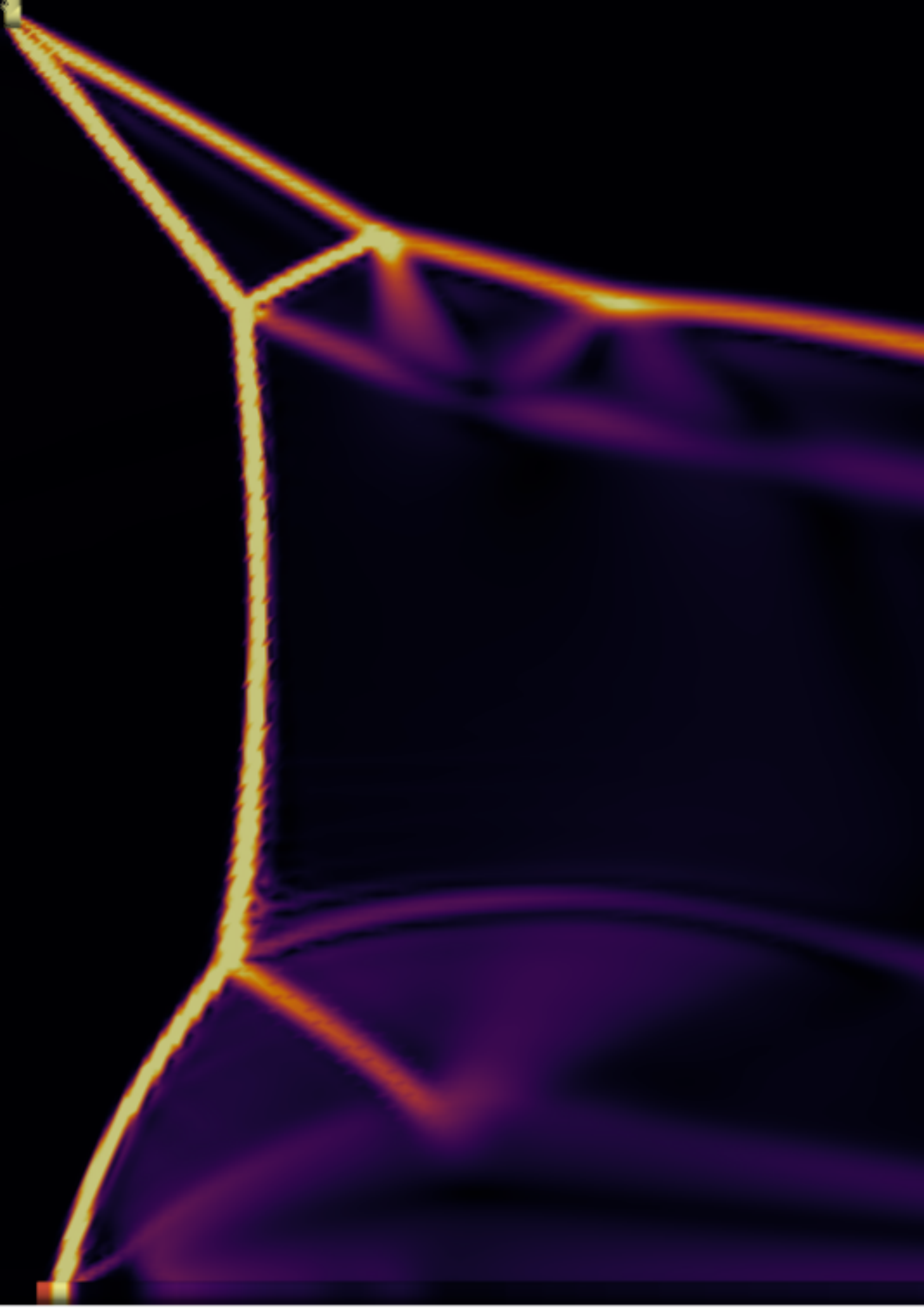}%
        \caption{\label{fig:reftype21}}
    \end{subfigure} 
    \begin{subfigure}{0.31\textwidth}
        \centering
        \includegraphics[height = 6cm,width =  \linewidth]{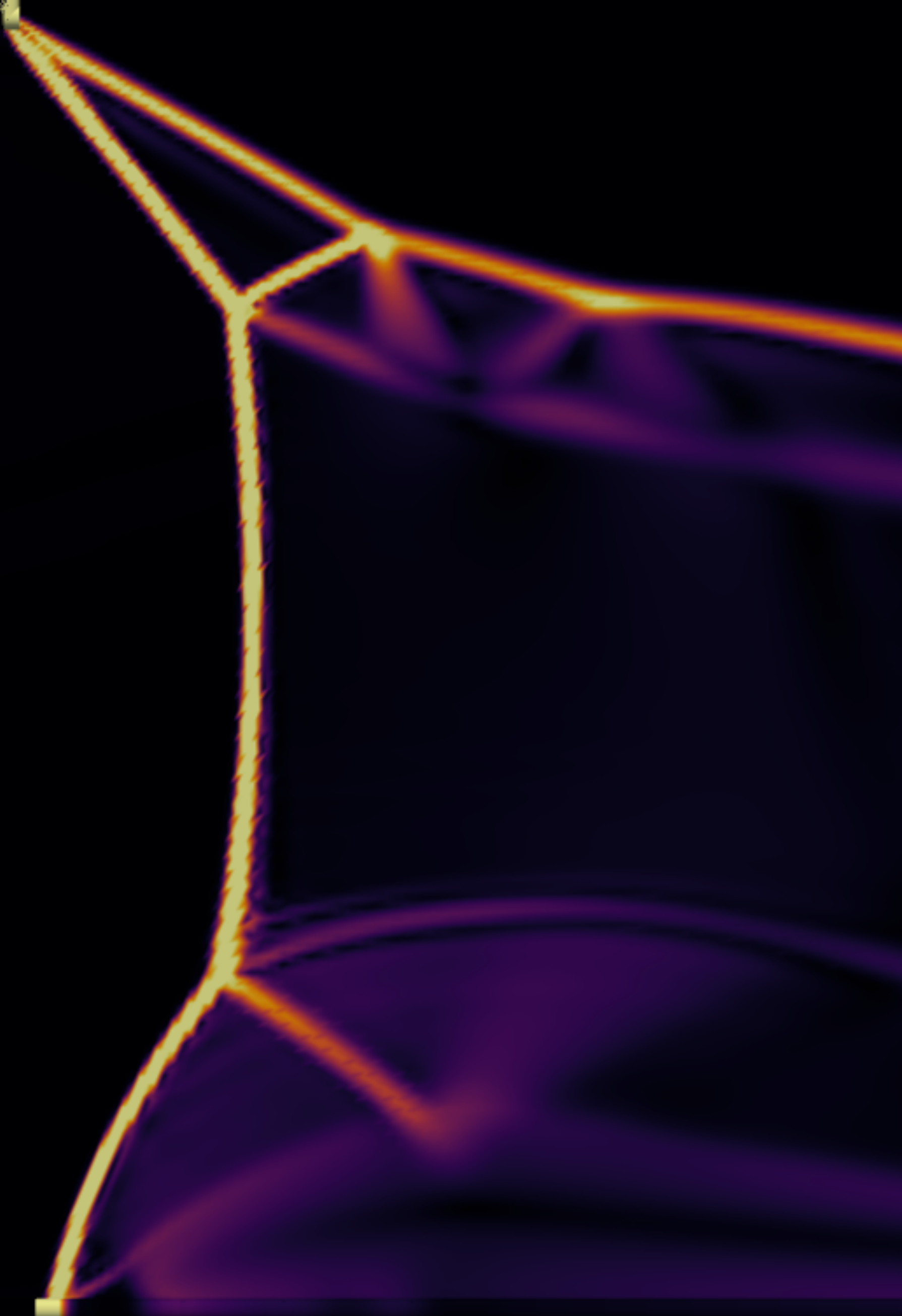}%
        \caption{\label{fig:reftype22}}
    \end{subfigure} 
    \begin{subfigure}{0.31\textwidth}
        \centering
        \includegraphics[height = 6cm, width =  \linewidth]{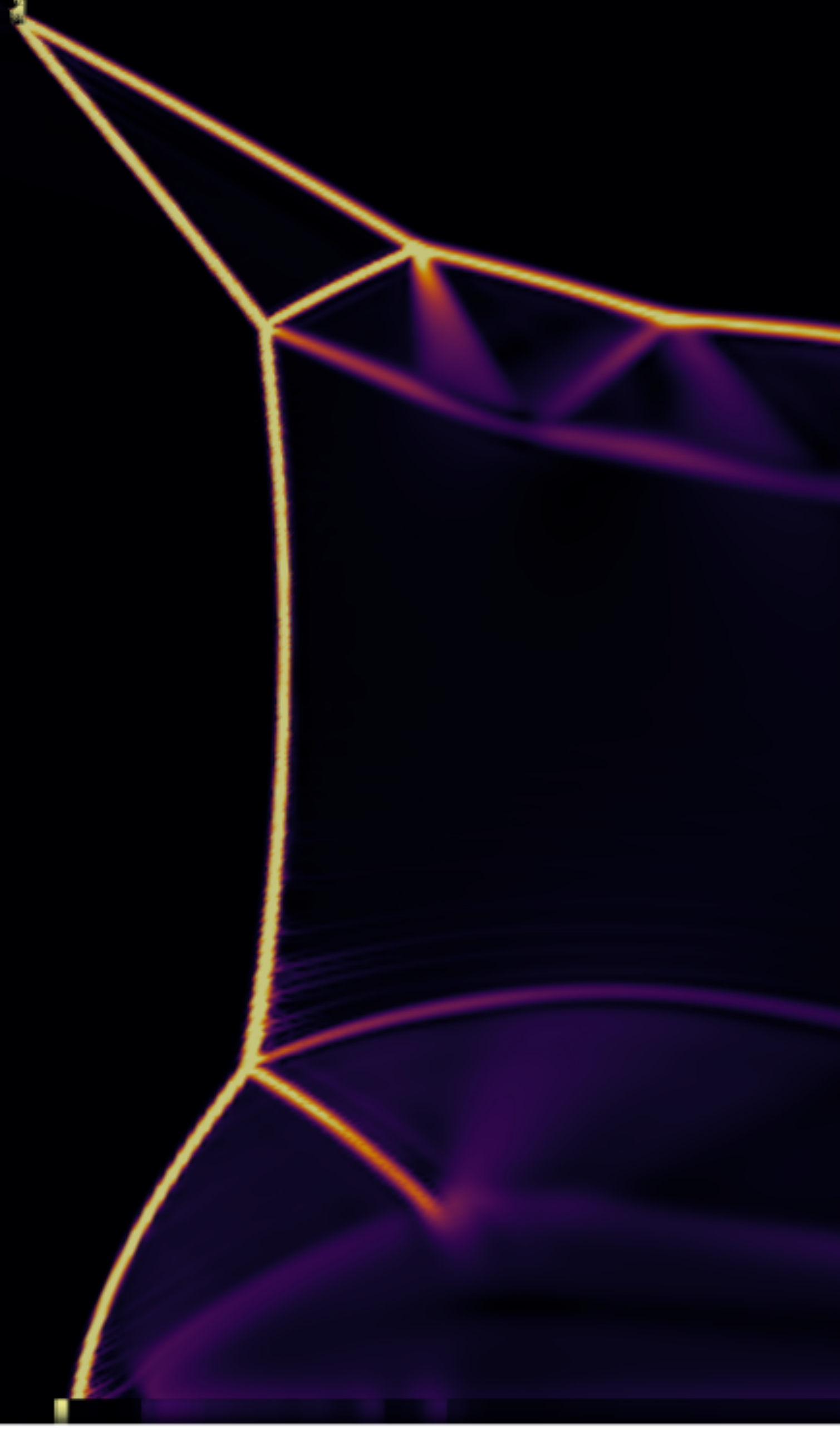}%
        \caption{\label{fig:reftype23}}
    \end{subfigure}
\caption{\label{fig:type20}Numerical schlieren images of the Type-II non-uniform csMR in the numerical simulations, with different grid densities and AMR levels (only the top half of the nozzle is shown). The three panels correspond to numerical resolutions at nozzle pressure ratio (NPR) $\approx 6.04$, (a) $550 \times 550$ cells, (b) $ 890 \times 890 $ cells, and (c) $1235 \times 1100$ cells.}
\end{figure}
An iso-contour of the \citet{lovely1999shock} shock-detection parameter is plotted in Fig.~\ref{fig:nice}. The salient features and global flow field organisation reveal a clear presence of the asymmetric Type-II cap-shock pattern, with four incident and reflected shocks altogether. A grid convergence study has also been conducted and the numerical schlieren images can be seen in Fig.~\ref{fig:type20}. We observe that much of the shock structure features remain the same for all grid resolutions. However, the coarser grids have slightly thicker shocks, while the finest mesh in Fig.~\ref{fig:type20} have significantly thinner ones. Nonetheless, the simulation demonstrates grid convergence and the finest mesh proves to achieve sufficient resolution. We therefore continue our study using the $1235\times 1100$ cell count case. Additional grid convergence studies have also been done for different configurations (Fig.~\ref{fig:regref}), where all results indicate that our solutions are accurate and independent of the grid resolution (see also Appendix~C for more extensive grid convergence studies).

\subsection{Symmetry case verification }
\begin{table}
\begin{tabular}{llllllllll}
     &                 &            & \multicolumn{7}{l}{$H_{mt}/H_t$}                                                                      \\
Case & $M_0$ & $\theta_1$ & Present & $w/H_t$ & Analytical$^1$ & Analytical$^2$ & Analytical$^3$ & Analytical$^4$ & Numerical$^3$ \\
1    & 2.84            & 20.8       & 0.1449  & 1.42    & 0.140          & 0.1175         & 0.202          & 0.118          & 0.191         \\
2    & 4.00            & 23.0       & 0.1437  & 1.28    & 0.106          & 0.0767         & 0.110          & 0.076          & 0.121         \\
3    & 4.00            & 25.0       & 0.2495  & 1.19    & 0.300          & 0.1704         & 0.213          & 0.167          & 0.223         \\
4    & 4.50            & 23.0       & 0.1295  & 1.10    & 0.051          & 0.0376         & 0.058          & 0.036          & 0.052         \\
5    & 4.96            & 28.0       & 0.3411  & 1.10    & 0.395          & 0.2742         & 0.292          & 0.269          & 0.283         \\
6    & 5.00            & 26.9       & 0.2981  & 1.10    & 0.296          & 0.1974         & 0.213          & 0.191          & 0.203        
\end{tabular}
  \caption{Comparison of the non-dimensional Mach stem height, $H_m/H$, with values from \citet{mouton2007mach}$^1$, \citet{Roy2019}$^2$, \citet{Gao2010}$^3$, \citet{Li1997}$^4$, for different wedge length ratios ($w/H_t$), all angles are in degrees. Note that none of these models has been derived with geometrical considerations of an over-expanded supersonic jet, and are for wedge-induced shocks.}
  \label{tab:coffee}
\end{table}
\subsubsection{Comparison of over-expanded jet with the wedge flows}
In the present analysis, we constrain our model to its symmetric version for purposes of verification with the available analytical, experimental and numerical data for the wedge flows, with different wedge length ratios ($w/H_t)$. Further to this  it was pointed out by \citet{paramanantham2022prediction} that it is important to acknowledge that comparisons between models for the over-expanded nozzle and the wedge flows is complicated by the fact that in the latter case, the Mach stem height depends on an additional length scale, $w$, the wedge length. This implies that the Mach stem height will vary for different $H/w$ ratios, while the over-expanded nozzle Mach stem height changes only with $H$. Therefore, it is expected that our results may deviate slightly from the wedge flows, which are derived and obtained based on different geometrical consideration. However, similar to the observations of \citet{li1998mach}, it will be shown that the Mach stem heights for both geometries are comparable, with all values still in good agreement for a wide range of $w/H$ ratios, which lends instrumental insight into the behaviour of overall Mach reflection configurations within internal flows. 

Also, the ability to adapt between asymmetric and symmetric methods is because our asymmetric extension of \citet{li1998mach}'s model is based on the ansatz of an initially non-uniform flow field, with different Mach numbers in the upper and lower domains. Therefore, the perfectly symmetric case just corresponds to the case where no non-uniform flow exists and the upstream Mach numbers are equal.
\begin{figure}
  \centering
\includegraphics[trim = 0cm 2cm 0cm 0cm, width = \linewidth]{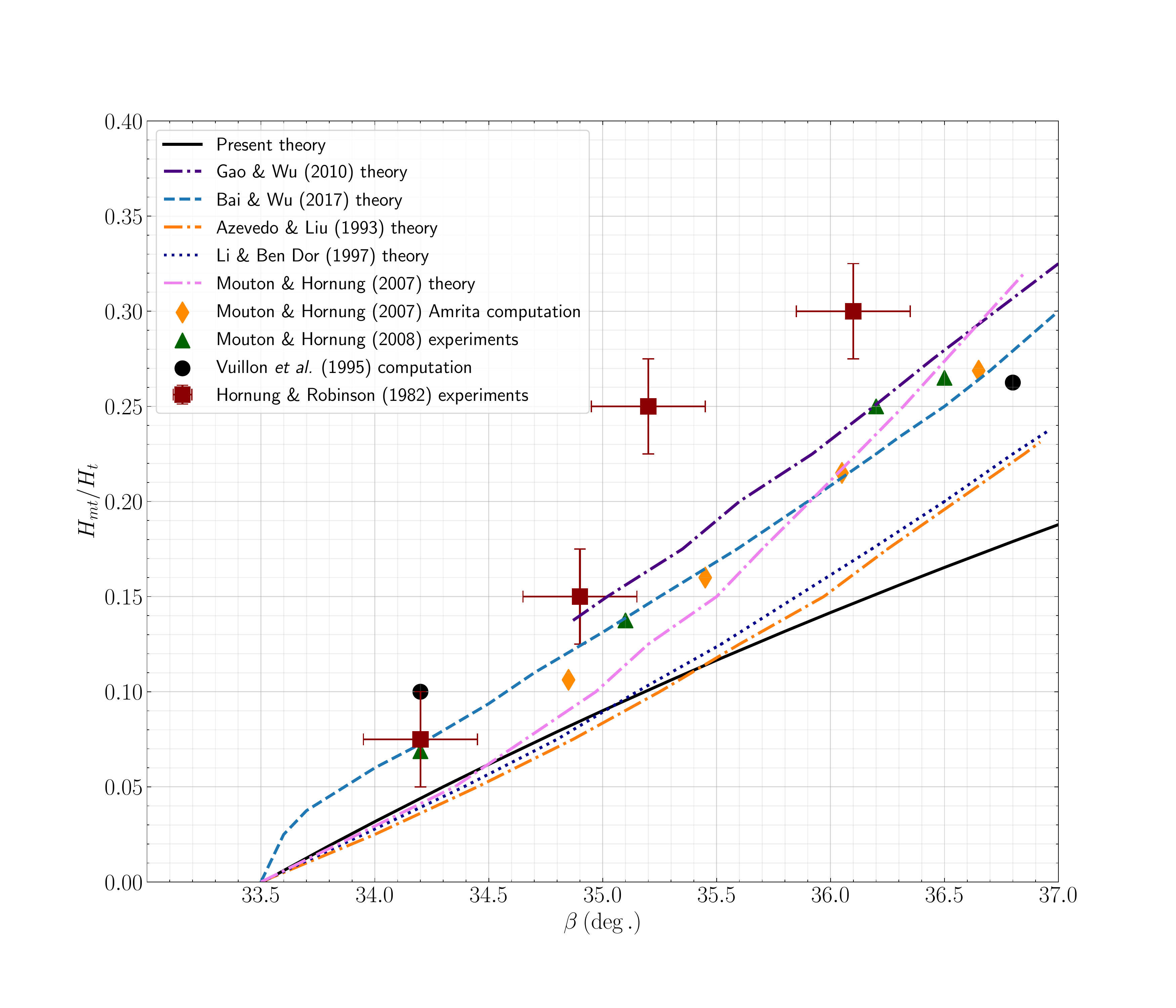}
  \caption{Comparisons of the non-dimensional Mach stem height, $H_{mt}/H_t$ against shock angle for $M = 3.98$, from current over-expanded jet theory with prior theories, experiments and computations for the wedge flows with the wedge length ratio, $H_t/w = 0.4$.}
\label{fig:first}
\end{figure}
\begin{figure}
    \begin{subfigure}{0.5\textwidth}
    \centering
    \includegraphics[ width = 1.1\linewidth]{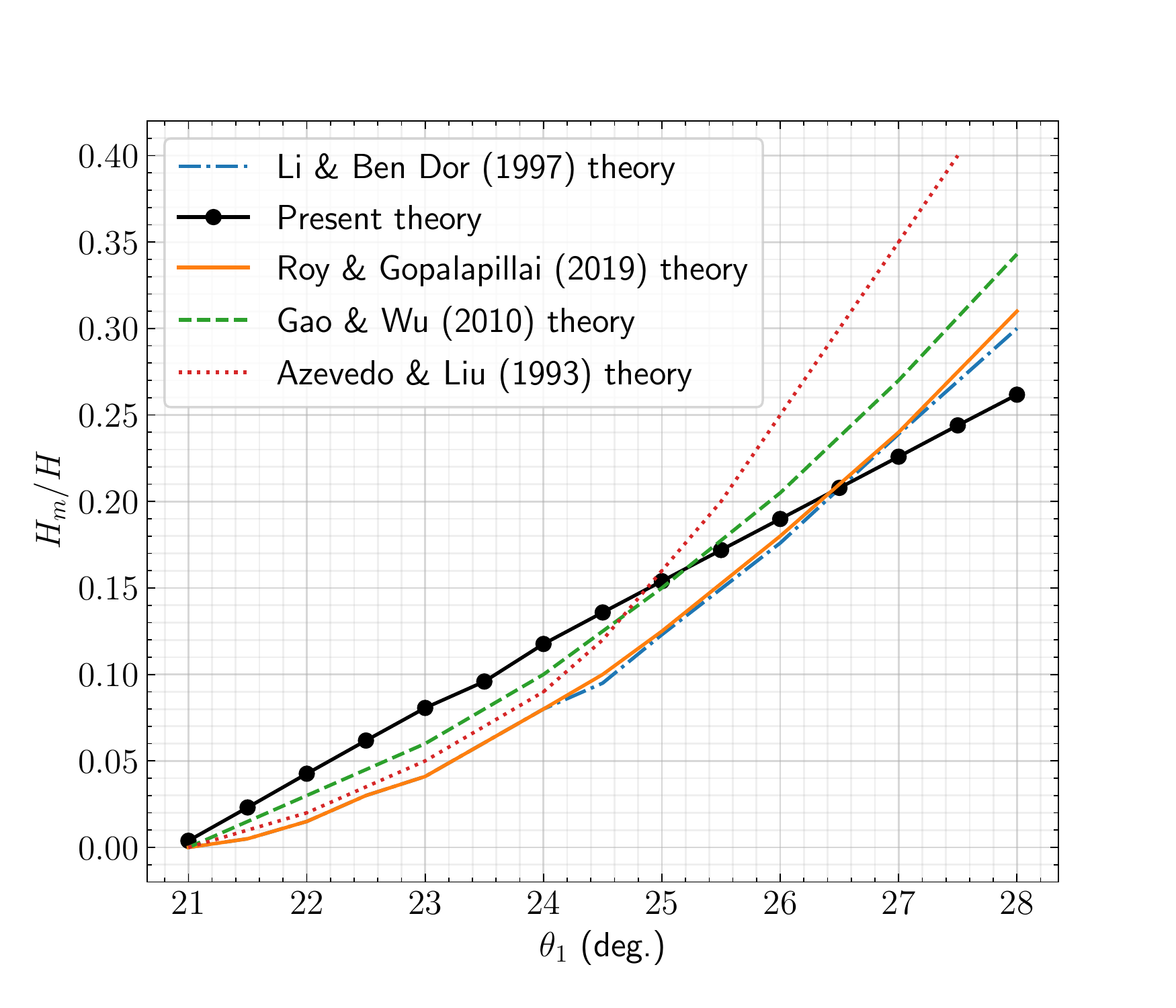}
    \caption{\label{fig:sym1}$M_0 = 4.5$}
    \end{subfigure}
    \begin{subfigure}{0.5\textwidth}
    \centering
    \includegraphics[width = 1.1\linewidth]{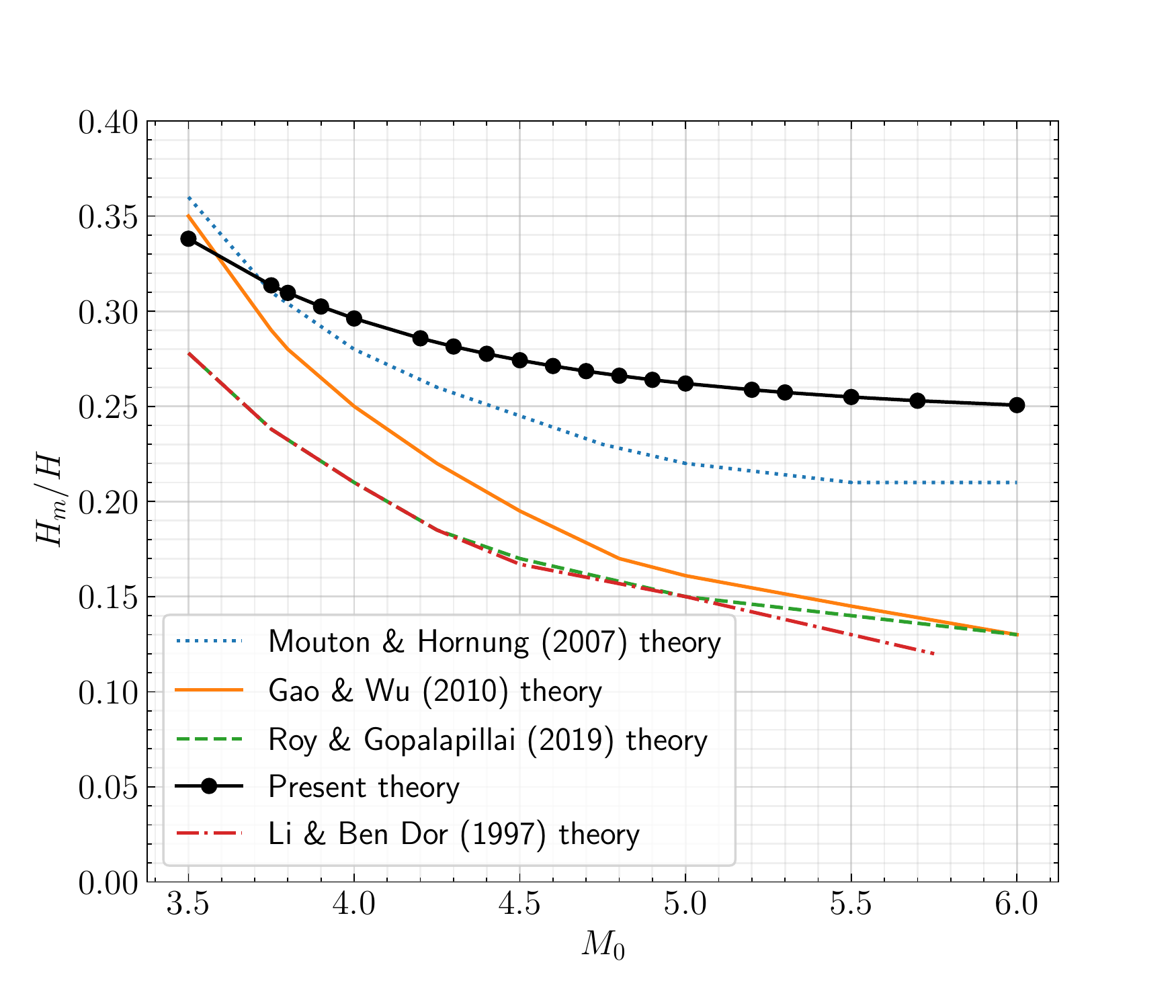}
    \caption{\label{fig:sym2}$\theta_1 = 26^{\circ}$}
    \end{subfigure}
\caption{\label{fig:symwedge}Comparisons of the non-dimensional Mach stem height, $H_m/H$ from present theory with those of \citet{Li1997}, \citet{Roy2019}, \citet{Azevedo1993} and \citet{Gao2010} and \citet{mouton2007mach}. Note that these theories have been developed for wedge flows, not over-expanded nozzle flows. The symmetric von Neumann criterion is predicted accurately at $\theta_1 = 21^{\circ}$ in Fig.\ref{fig:sym1}. }
\end{figure}

\begin{figure}
 \centering
\includegraphics[trim = 0cm 2cm 1cm 0cm,width = \linewidth]{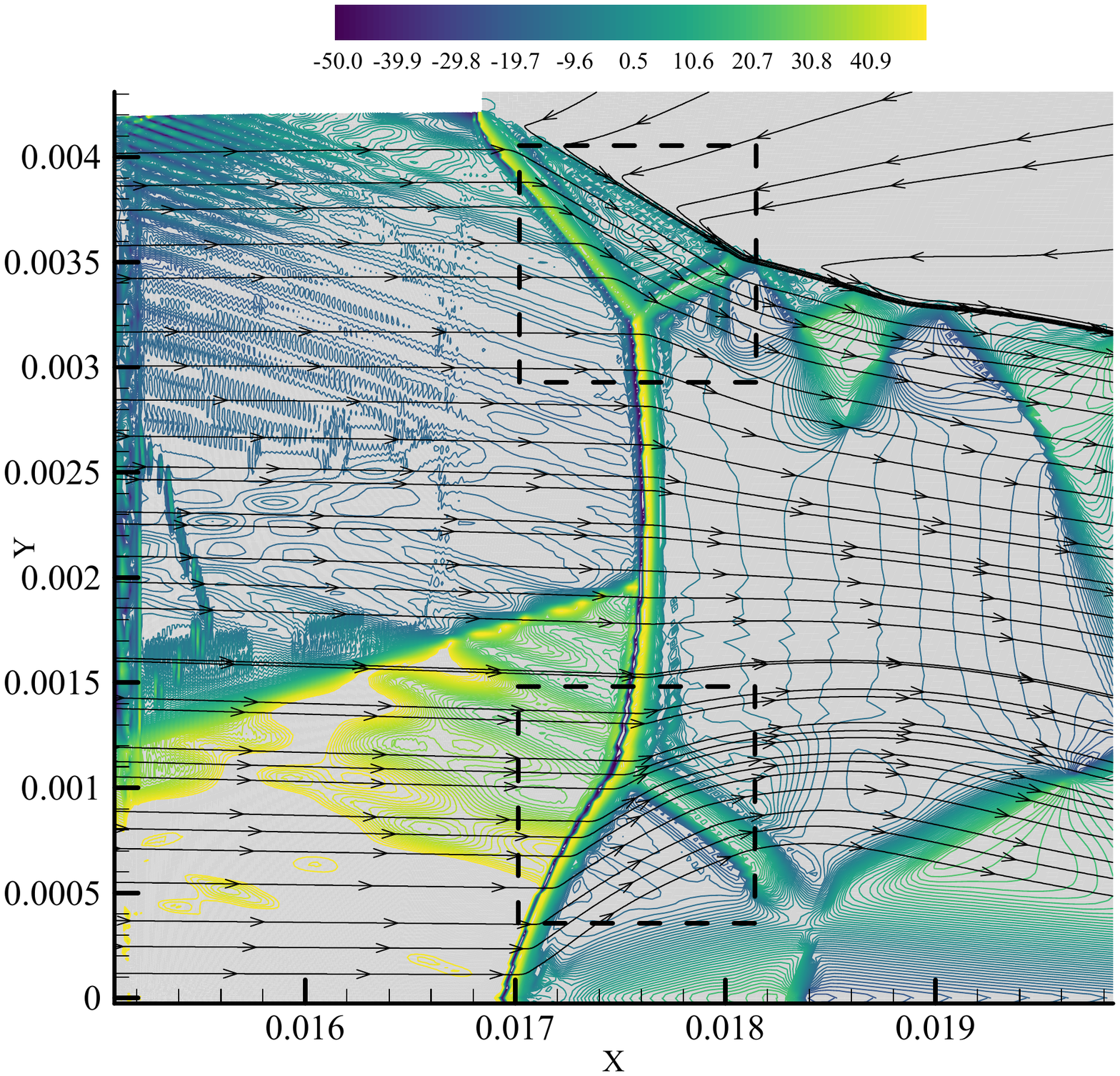}
  \caption{Close-up view of Fig. \ref{fig:nice} of the Type-II csMR configuration, where only the top half of the nozzle is shown with flow streamlines, showing the deflections across the shocks.}
\label{fig:close}
\end{figure}

The comparison of non-dimensional Mach stem height against incident shock angle is shown in Fig.~\ref{fig:first}, where it can be readily observed that our model follows the same trend as the other models, numerical computations and experiments for the wedge flows with increasing shock angle \citep{mouton2007mach,Mouton2008,Gao2010,Bai2017,Azevedo1993,vuillon1995reconsideration,hornung1982transition}. Of note as well is the fact that the Mach stem height in our model drops to zero at $\beta \approx 31\deg$, which corresponds exactly to the symmetric von Neumann criterion at $M_0 = 3.98$, so the present model accurately captures the transition to RR at the specified shock angle and Mach number. Since we previously noted that the StMR-StMR configuration occurs exactly at the von Neumann criterion, we do not observe this structure in the steady flow regime due to a transition to oRR, where the Mach stem height reduces to zero. It is therefore possible to comment on this analysis that while the theoretical StMR-StMR configuration is never reached in this case, the same can not be said for any future models derived for pseudosteady or unsteady flows, owing to the fact that the Mach stem height should reduce to zero at the detachment or sonic criterion in those cases. 

The remainder of the figures here, namely the two plots in Fig.~\ref{fig:symwedge}, similarly demonstrates that our model is in good agreement with other known models, despite the fact that they have been developed for very different geometries. This indicates that our method can also be applied for parametric analyses of wedge-induced shock reflections. However, as mentioned earlier, comparisons between models for the over-expanded nozzle and the wedge flows is complicated by the fact that in the latter case, the Mach stem height depends on an additional length scale, $w$, the wedge length. But similar to \citet{li1998mach}, our results indicate that the Mach stem heights for both geometries are comparable, with all values still in good agreement for a wide range of $w/H$ ratios, as seen in Table~\ref{tab:coffee} and all the figures discussed.

\subsubsection{Variation of the Mach stem height}
To further comment on this analysis, the algorithm employed shows that the symmetric von Neumann criterion is well-captured (Figs.\ref{fig:first} and \ref{fig:sym1}), where at a certain deflection angle, the Mach stem height decreases to zero, corresponding to transition to oRR (overall regular reflection). Moreover, no dynamic effects \citep{naidoo2014dynamic} are observed as well within the present model. 

In comparing the present analytical model with those for the wedge flows for different wedge length ratios (Table~\ref{tab:coffee}), the Mach stem height in all cases are in reasonable agreement with them, and fall in between the models of \citet{Gao2010} and \citet{mouton2007mach}. However, it can be seen that the model predicts a higher value of the Mach stem height with increasing flow deflection angle; and correspondingly decreasing shock angle (Fig.~\ref{fig:first}), the same conclusion reached by \citet*{paramanantham2022prediction}. Moreover, the variation of the Mach stem height (slope of the $H_m/H_t$ curve) decreases at higher flow deflection and inflow Mach numbers when compared to the wedge flows, as seen in Figs.~\ref{fig:first} and~\ref{fig:symwedge}. We believe this attribute suggests that the stability of the Mach reflection inside an over-expanded nozzle does not decline as rapidly as with the flow between two wedges, elucidating the idea that MR structures are more likely to occur and remain in an over-expanded jet, with the Mach stem height not changing significantly up to the point where the shock is so strong that reflection cannot occur ($\theta \geq \theta_{max})$ \citep{Gao2010}.

\subsubsection{Comparison with over-expanded jet flows}
\begin{figure}
 \centering
\includegraphics[width = 0.8\linewidth]{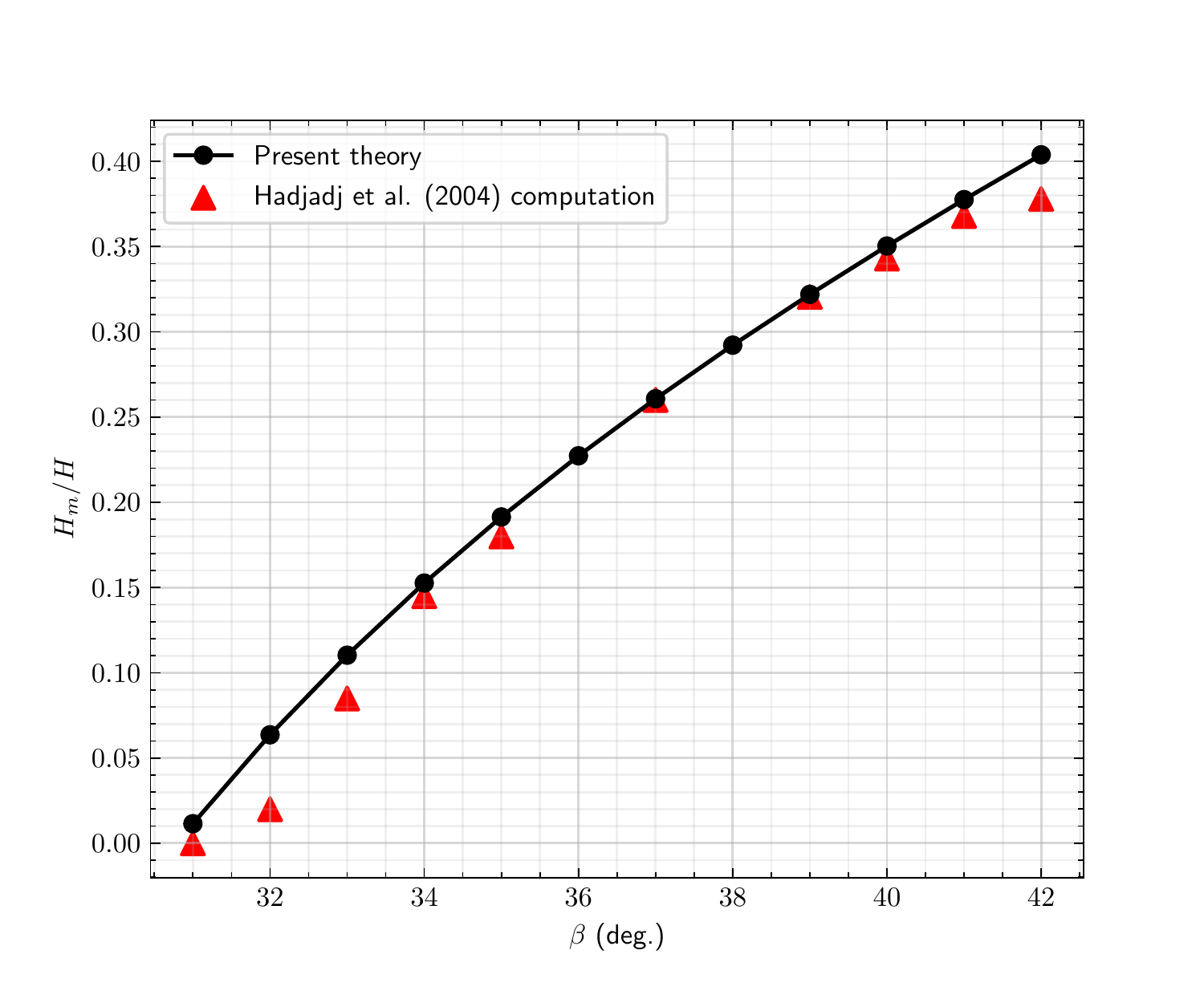}
  \caption{Comparisons of the non-dimensional Mach stem height with computational fluid dynamics (CFD) simulations by \citet{hadjad2004}, for their over-expanded nozzle geometry at $M_0 = 5$.}
\label{fig:hadjadj}
\end{figure}
\begin{figure}
 \centering
\includegraphics[width = 0.8\linewidth]{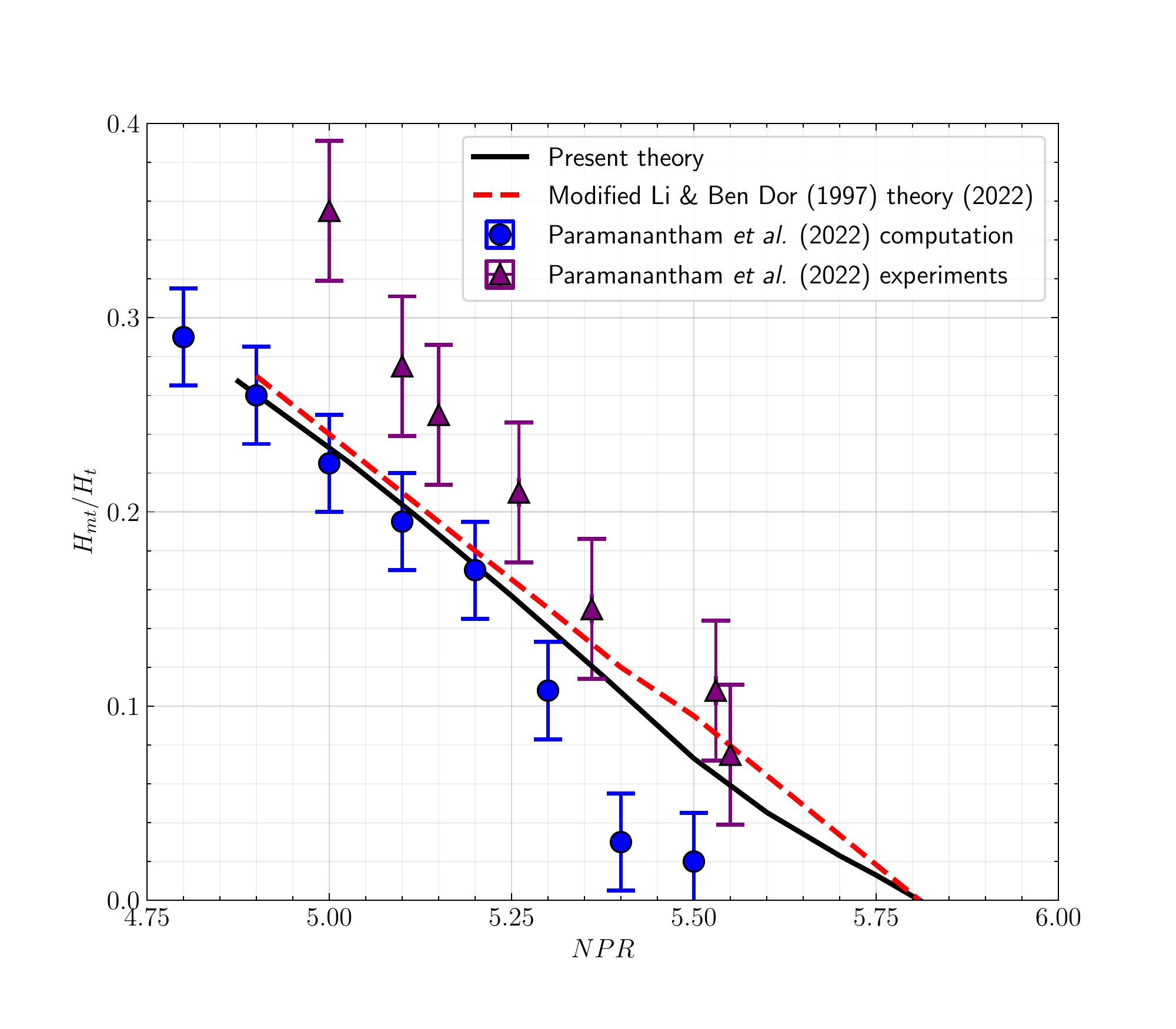}
  \caption{Comparison of the non-dimensional Mach stem height for the over-expanded jet from present theory with the computation and experiments by \citet{paramanantham2022prediction}, and their modified \citet{Li1997} for the open jet flow at $M_0 = 2.44$.}
\label{fig:jetparamantham}
\end{figure}
The calculated non-dimensional Mach stem heights from in the present model are now validated with numerical, analytical and experimental values from prior works on over-expanded jet flows. We compare our values against the computations of \citet{hadjad2004}, for their over-expanded nozzle at $M_0 = 5$ (Fig. \ref{fig:hadjadj}). It can be readily seen that our model is in excellent agreement with numerical results specific to nozzle flows, provided that the hysteresis loops in their data are neglected, which cannot be accounted for by any analytical models as of yet. The excellent agreement in the Mach stem heights are consistent with their observation that viscous effects do not generally play a significant role in the morphology of the shock structure and their transitions.

While Mach stem height data for the over-expanded nozzle geometries are somewhat lacking in the literature, the numerical data from \citet{hadjad2004}, \citet*{paramanantham2022prediction}, and experimental data from their schlieren measurements (Fig.~\ref{fig:jetparamantham}), all demonstrate that our model matches closely with the numerical results. Fig.~\ref{fig:jetparamantham} further indicates that the present model closely matches the modified \citet{Li1997} method up until the von Neumann condition, but at the same time the present model follows the numerical results from \citet*{paramanantham2022prediction} more accurately.

Regarding the deviations between the computations and analytical models from experiments in Fig.~\ref{fig:jetparamantham}, \citet*{paramanantham2022prediction} suggested that the over-predictions in all cases may be due to viscous forces and/or shock-wave/turbulent boundary layer interactions, which alter the shock strengths and therefore potentially the appearance of the MR configuration. It can also be attributed to three-dimensional edge effects \citep{skews2000three}, due to the lateral expansion of the jet in the experiment, which is known to either accelerate or delay MR~$\leftrightarrow$~RR transition \citep{skews2000three,surujhlal2019three} and contaminate the comparison to two-dimensional MR models.  Nevertheless, the closer agreement between the analytical model and computational results suggests that our model is highly accurate in predicting the Mach stem heights for symmetric MR configurations. 

\begin{figure}
 \centering
\includegraphics[width = 0.8\linewidth]{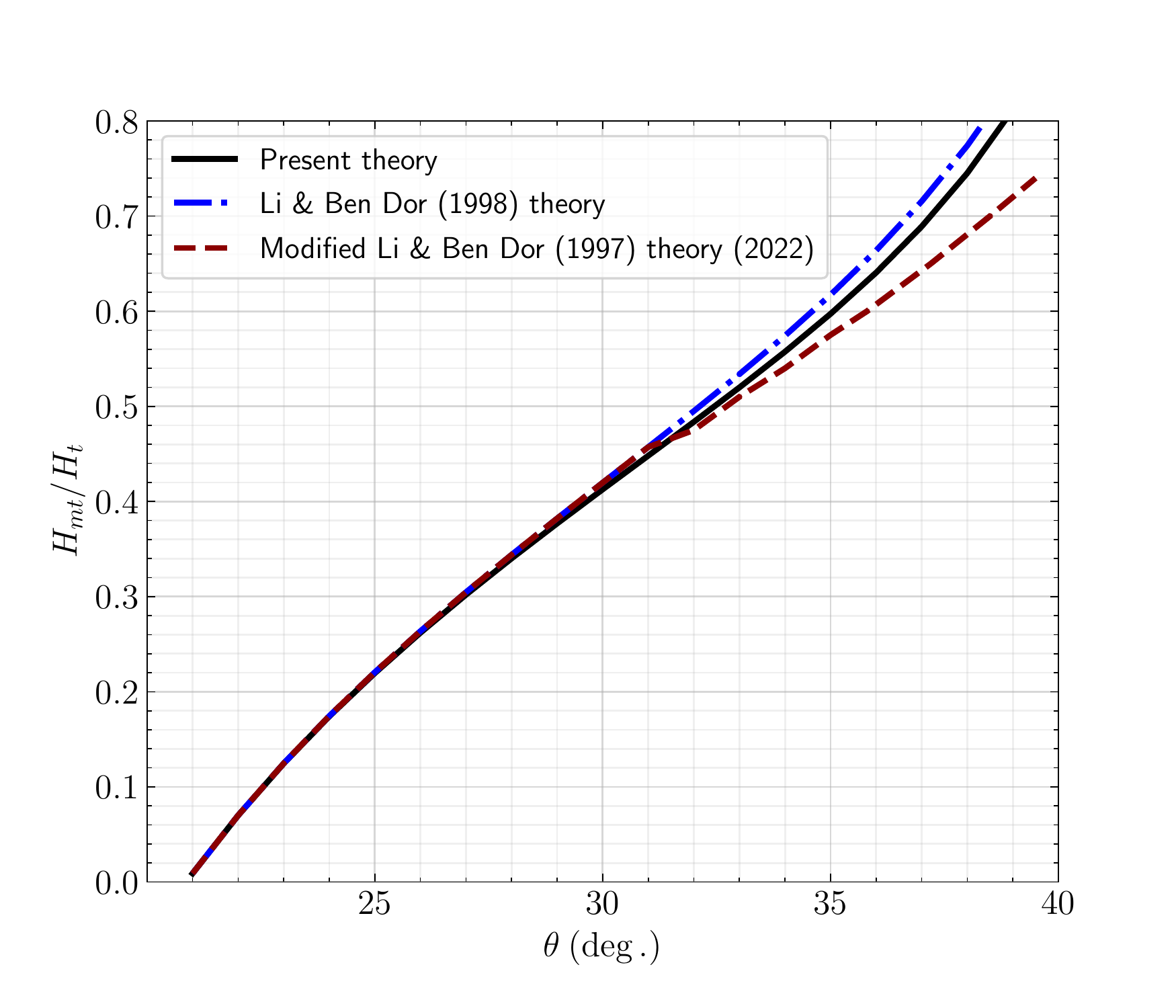}
  \caption{Comparison of the non-dimensional Mach stem height for the over-expanded jet from present theory with the theory of \citet{li1998mach} and the modified \citet{Li1997} by \citet*{paramanantham2022prediction} for $M_0 = 5$.}
\label{fig:libendorcompare}
\end{figure}

In order to further elucidate the differences between the present model and that of \citet{li1998mach} and the modified \citet{Li1997} method from \citet*{paramanantham2022prediction}, we plot the Mach stem heights against flow deflection angle for $M_0 = 5$ (Fig.~\ref{fig:libendorcompare}), where the present theory seems to fall in between both methods. The deviations between the present model and \citet{li1998mach} are due to the inclusion of a third expansion fan at point J (Fig.~\ref{fig:crazy}), as well as the consideration of averaged flow deflection angles \citep{Tao2017}, which retains a small but non-zero value in a perfectly symmetric MR configuration. However, the disparity is minute and seems to result in excellent agreement with relevant numerical and analytical data. It has also been separately verified that once those changes have been removed, we retain the same results as that of \citet{li1998mach}.

\subsection{Asymmetric oMR configurations}
To study the properties of asymmetric oMR configurations, we perform calculations of the Mach stem height for the uniform upstream Mach number $M_0 = 4.96$, so as to validate the results with transition lines considered by \citet{li1999analytical} and \citet{ivanov2002}. The upper flow deflection angle is fixed at $\theta_{1u}$ of $22^{\circ}$, while the lower flow deflection angle is reduced until the Mach stem height approaches null at the asymmetric von Neumann condition at $19.65^{\circ}$.
\begin{figure}
    \begin{subfigure}{0.5\textwidth}
    \centering
    \includegraphics[ width = 1.1\linewidth]{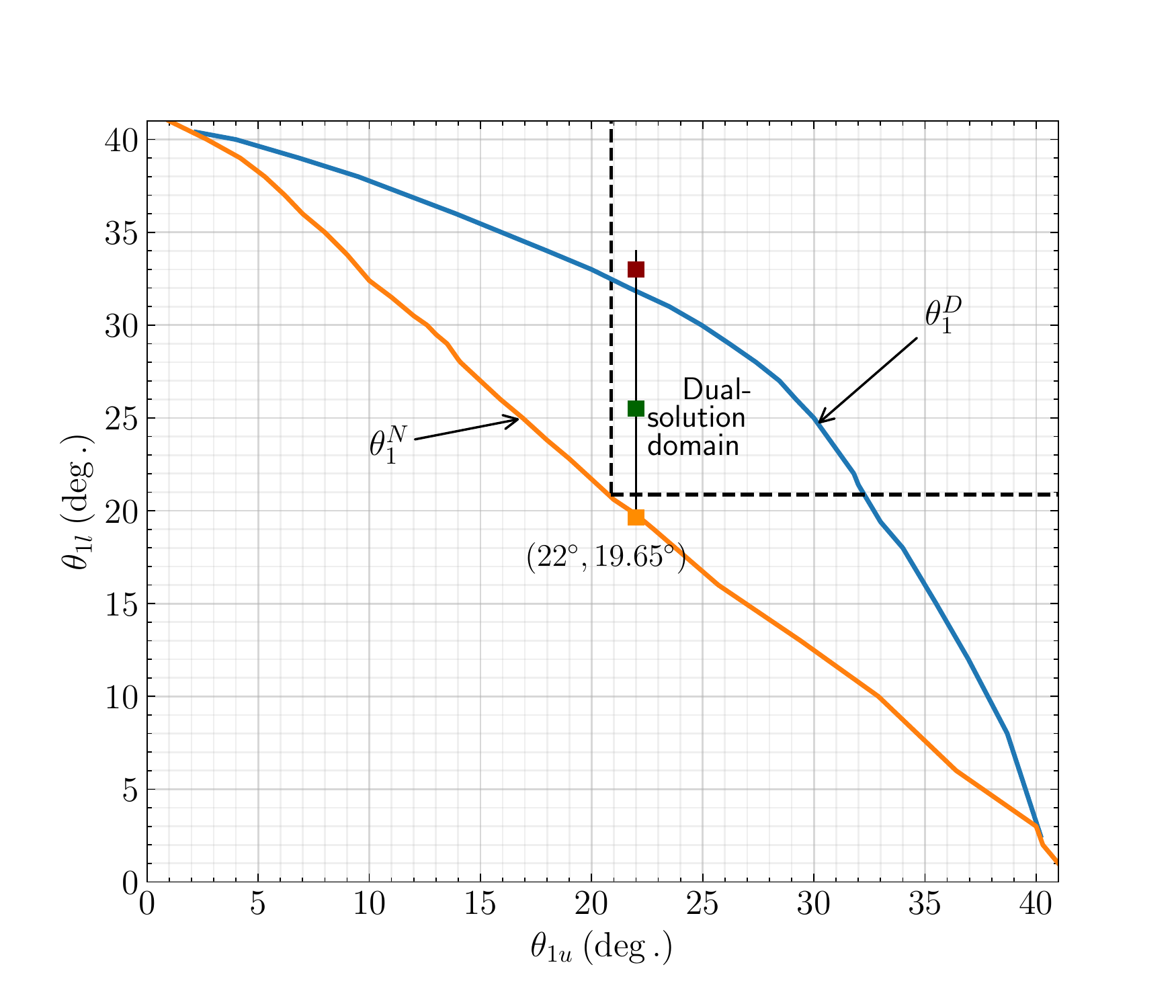}
    \caption{\label{fig:dual1}}
    \end{subfigure}
    \begin{subfigure}{0.5\textwidth}
    \centering
    \includegraphics[width = 1.1\linewidth]{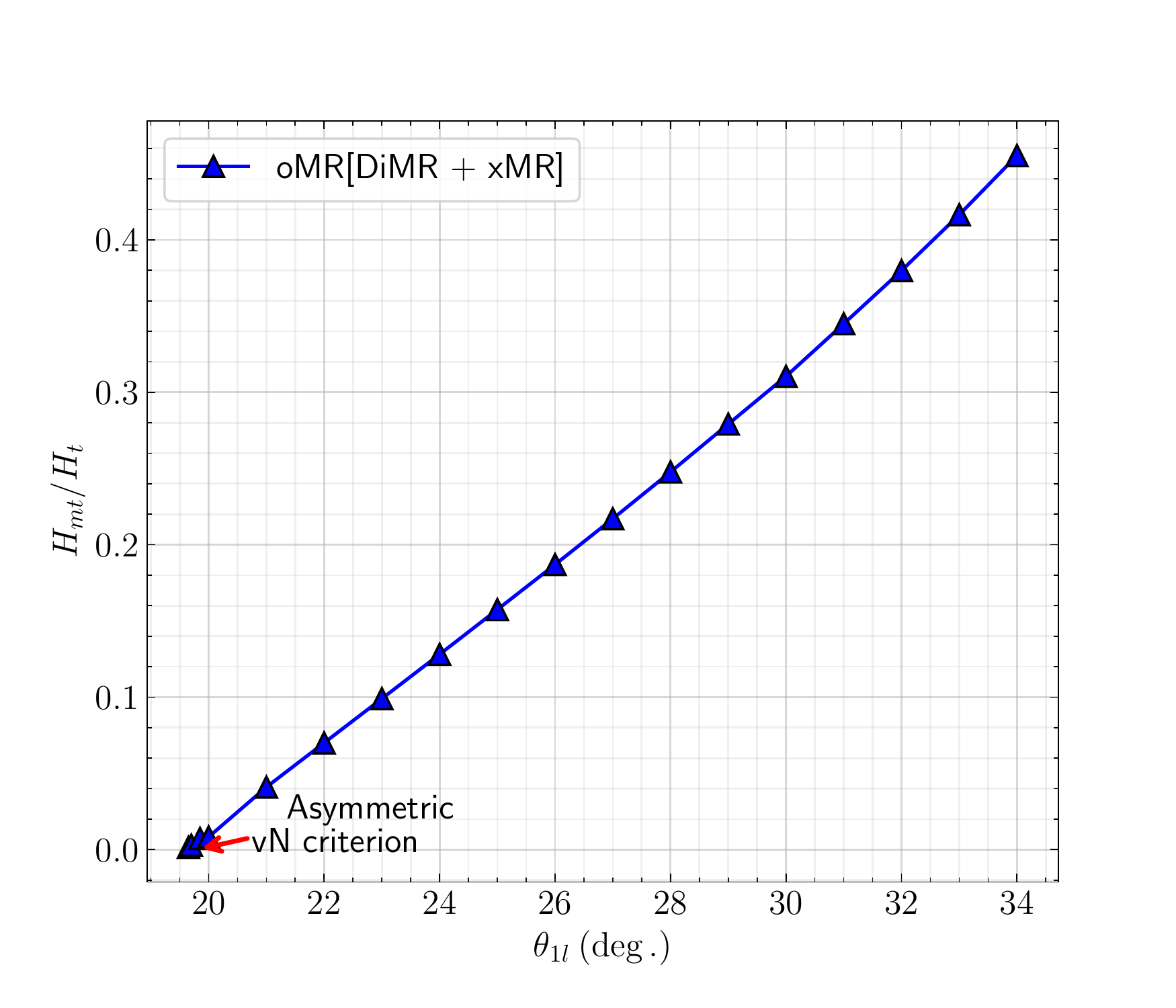}
    \caption{\label{fig:dual2}}
    \end{subfigure}
    \begin{subfigure}{\textwidth}
    \centering
    \includegraphics[width = 0.6\linewidth]{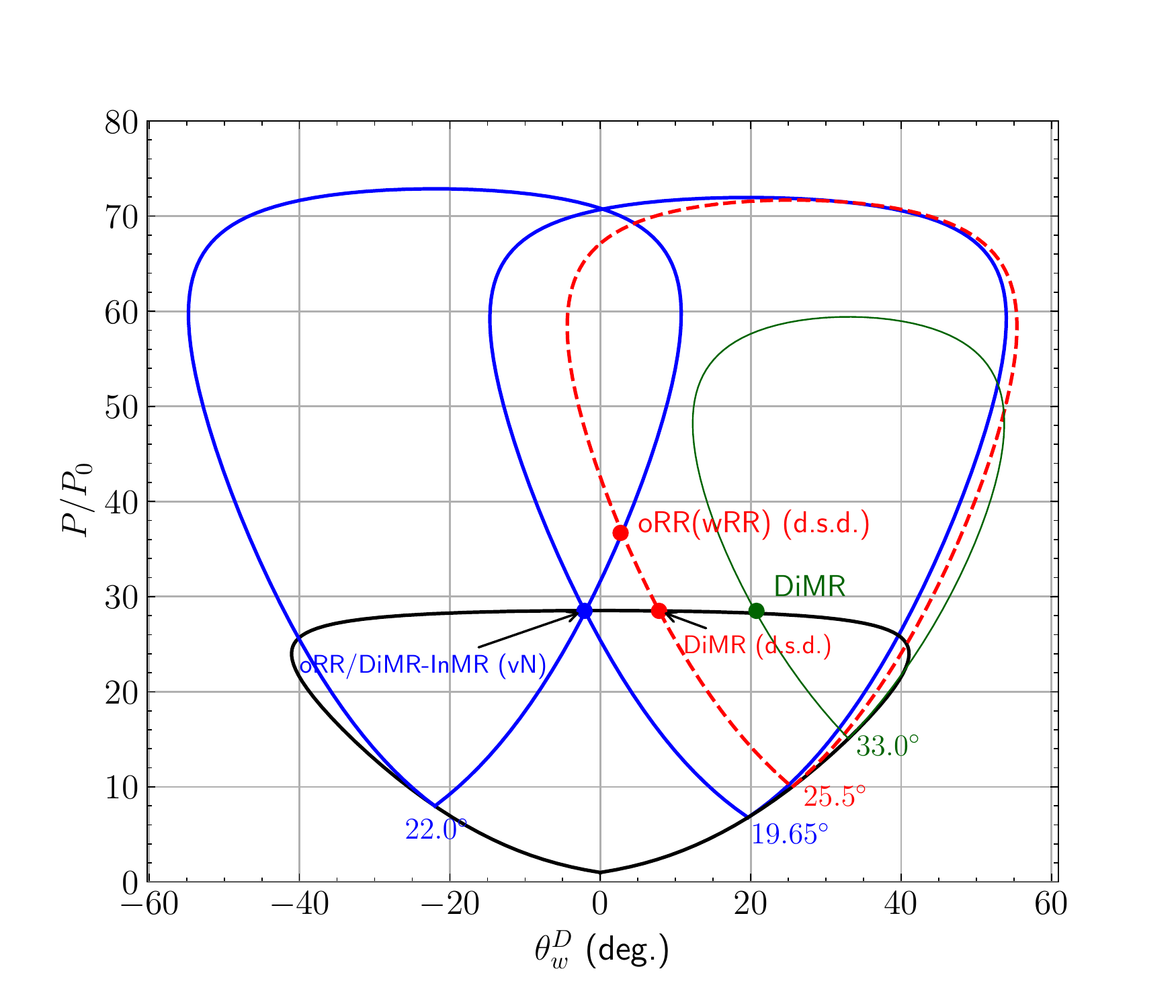}
    \caption{\label{fig:dual3}}
    \end{subfigure}
    
\caption{\label{fig:asymvn} Asymmetric von Neumann criterion for $M_0 = 4.96$ (a) Dual-solution domain in the ($\theta_{1u},\theta_{1l}$)- plane \citep{li1999analytical,ivanov2002}; (b) Plot of the normalised Mach stem height against lower flow deflection angle $\theta_{1l}$ for oMR[DiMR+DiMR] and oMR[DiMR+InMR] (abbreviated as oMR[DiMR + xMR]); (c) Shock polar representations of some of the oMR[DiMR+DiMR] and oMR[DiMR+InMR] cases. }
\end{figure}
The transition line taken is shown in Fig.~\ref{fig:dual1}, with the dual-solution domain plot from \citet{ivanov2002} in the $(\theta_{1u},\theta_{1l})$ plane, and the computed Mach stem heights are shown in Fig.~\ref{fig:dual2}. It can be observed that the model accurately predicts the asymmetric von Neumann criterion at $\theta_{1l}$ of $19.65^{\circ}$, as verified by the shock-polar theory cases shown in Fig,\ref{fig:dual3}, where the I-polar and both R-polars intersects exactly at the calculated von Neumann condition for the tested oMR[DiMR + InMR] configuration.

\subsection{Comparisons to the numerical simulations in the asymmetry case}
By incorporating asymmetry through the introduction of the non-uniformity in the upstream Mach number, we utilise the model prediction to compare it with the non-uniform Type-II Mach reflection \citep{Hadjadj2009,Nasuti2009} we observe in our numerical simulations. 

\begin{figure}
  \centering
\includegraphics[width = \linewidth]{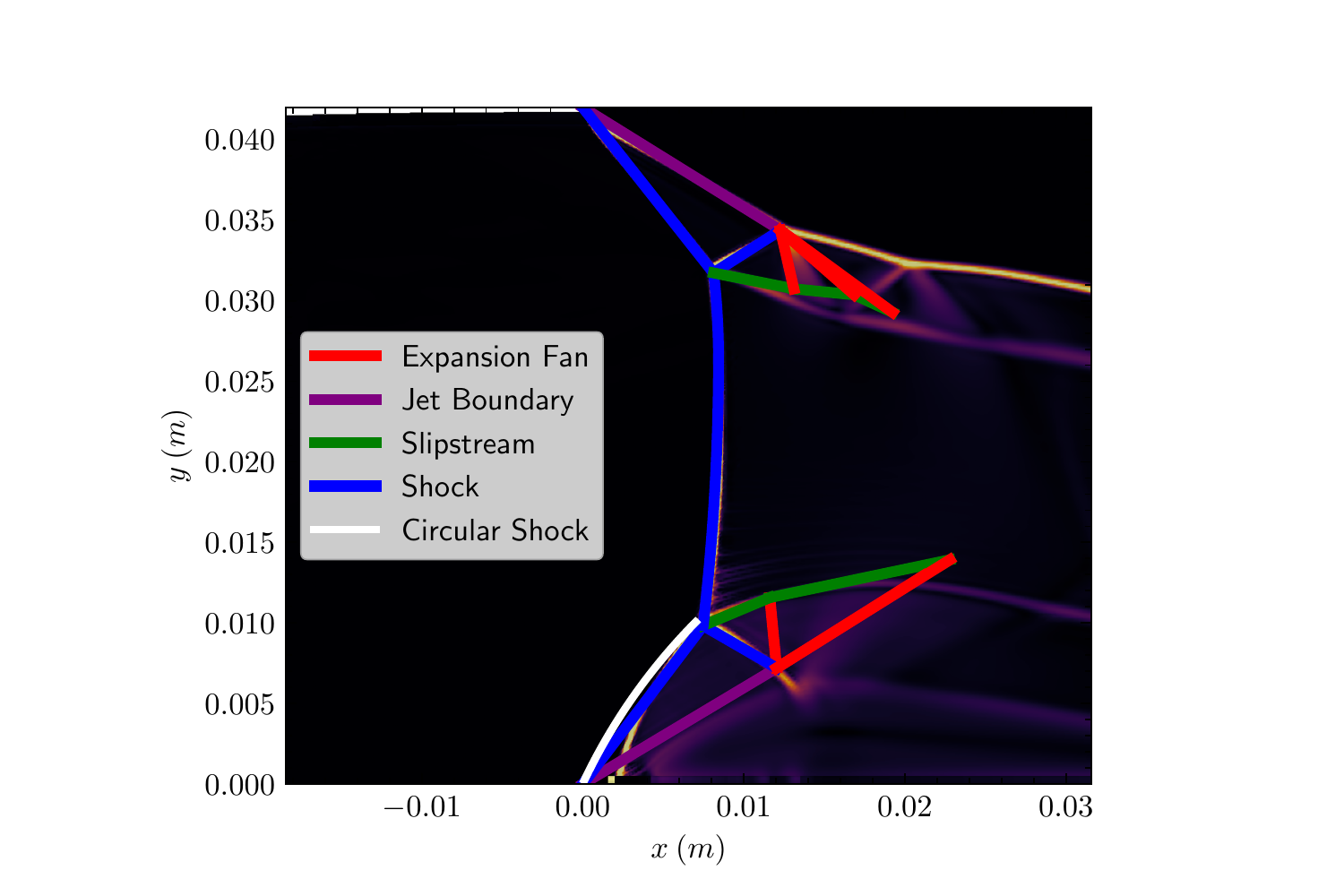}
  \caption{Comparisons of the Type-II cap-shock Mach reflection from our analytical model with our numerical simulations, $(M_{0u},M_{0l})= (2.80,3.27)$. The colour outline is the analytical model, while the background colour map shows the results of the numerical simulation.}
\label{fig:check}
\end{figure}

It can be seen that for the non-uniform Mach number pair $(M_{0u},M_{0l}) = (2.80, 3.27)$ (Fig.~\ref{fig:check}), the model adequately predicts all salient features of the shock structure in the first shock cell, including the Mach stem height and curvatures, wherein the circular-type Mach stem was approximated as a circular arc in the osculating plane, and the other as a second-order curve of \citet{Li1997}; the oblique shock-fan-jet boundary interaction, where the expansion fans then emanate downward to the slipstream, inducing changes in its orientation. Therefore, we find excellent agreement between the analytical model and the numerical simulation.

\subsection{Curvature of the central Mach stem (MS$_1$)}
\begin{figure}
    \begin{subfigure}{0.5\textwidth}
    \centering
    \includegraphics[ width = \linewidth]{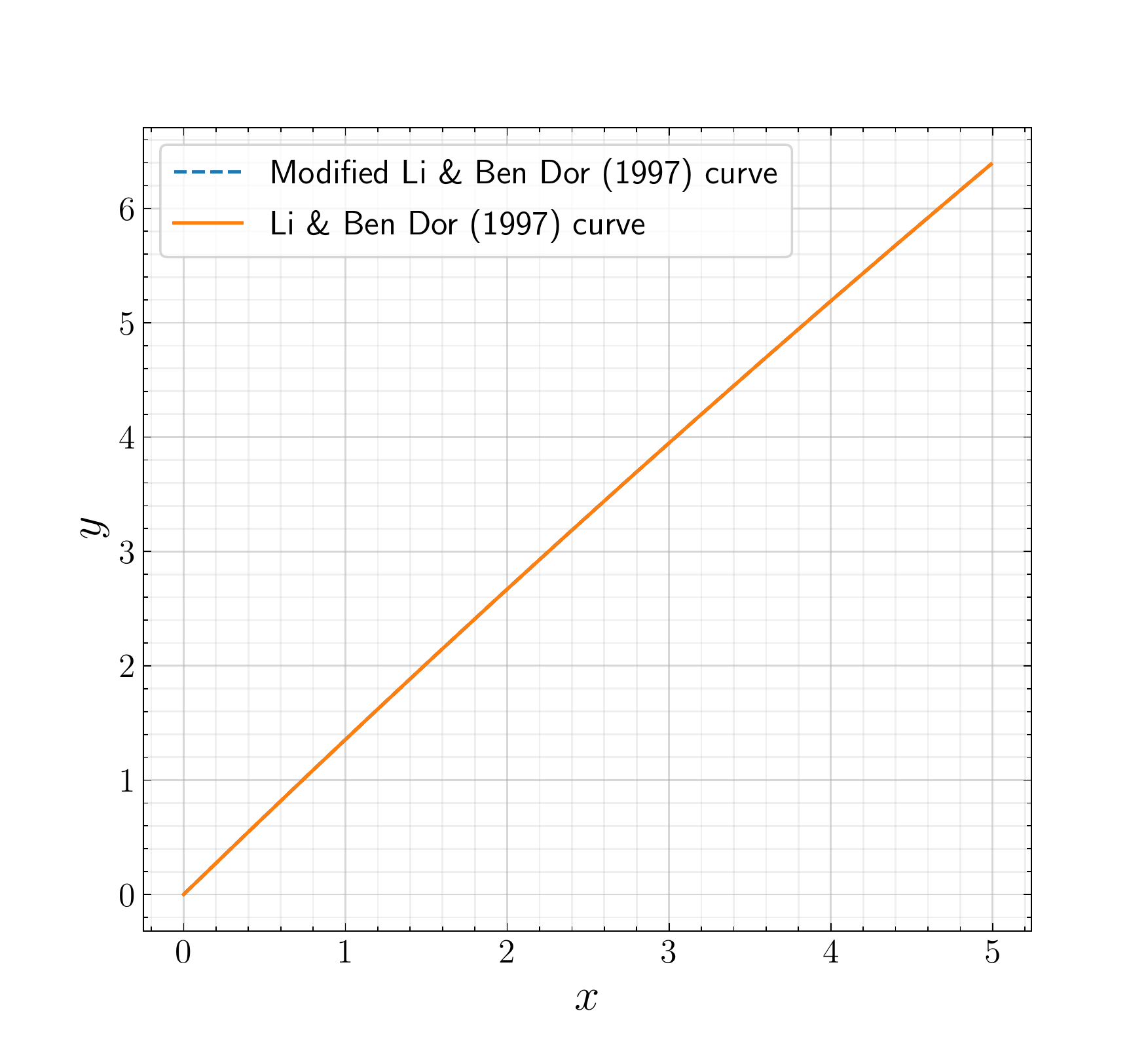}
    \caption{\label{fig:libendor}}
    \end{subfigure}
    \begin{subfigure}{0.5\textwidth}
    \centering
    \includegraphics[width = \linewidth]{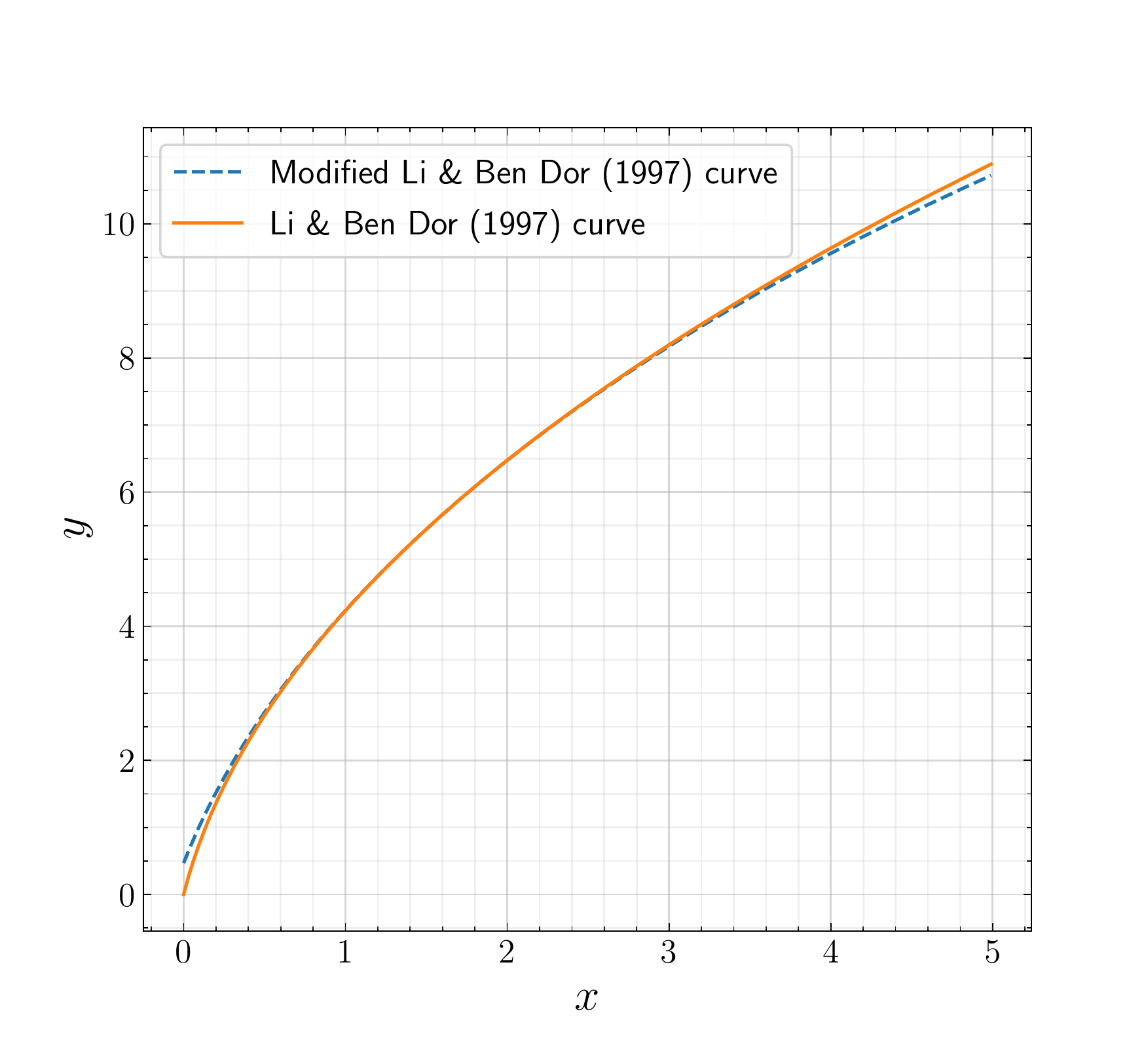}
    \caption{\label{fig:libendor1}}
    \end{subfigure}
    \begin{subfigure}{\textwidth}
    \centering
    \includegraphics[width = 0.6\linewidth]{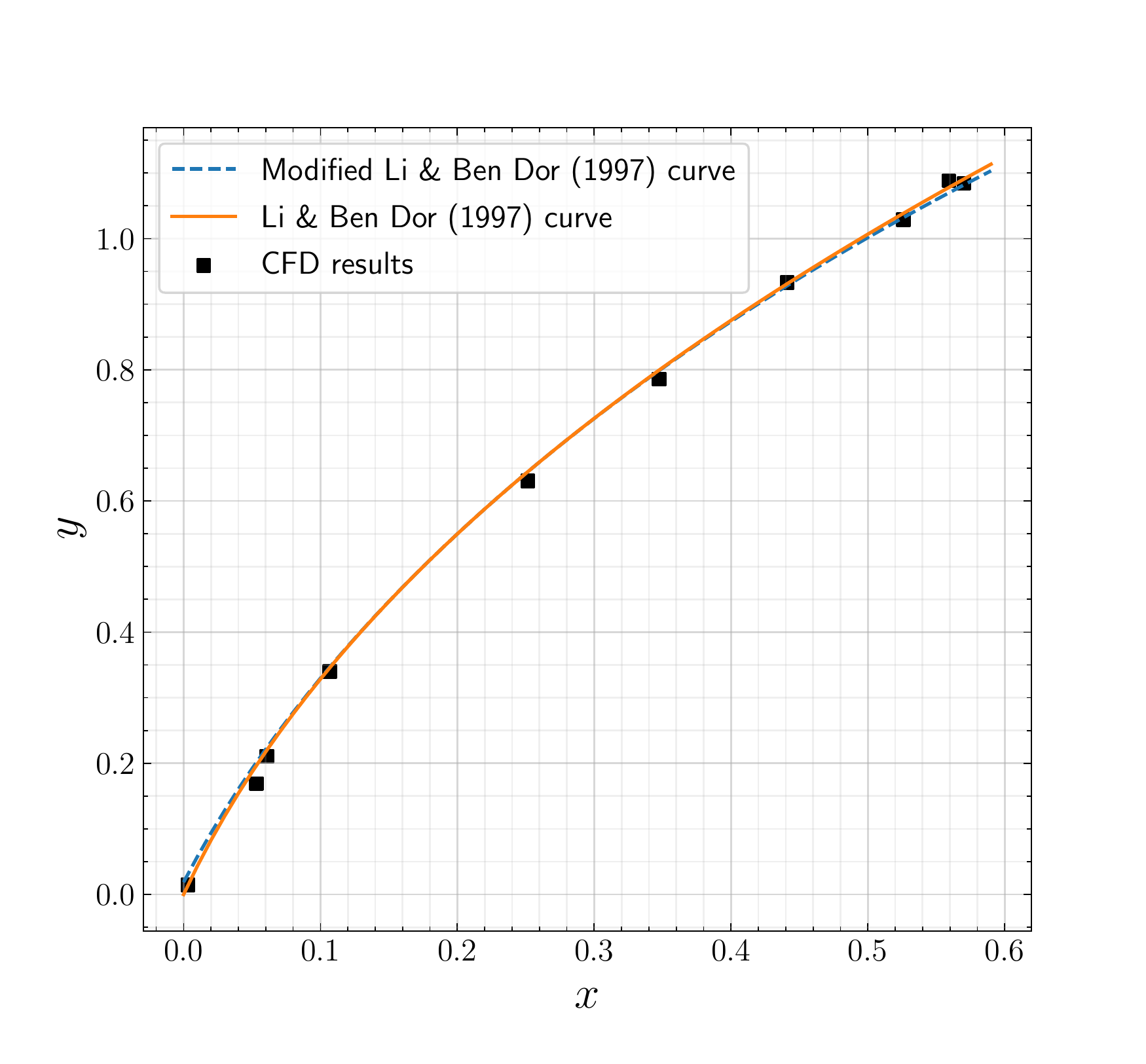}
    \caption{\label{fig:libendornumerical}}
    \end{subfigure}
    
\caption{\label{fig:libendor3} Comparison of the \citet{Li1997} second order curve with present curve based on osculating circle approximation for various slope angles, and with CFD results for curvature of central Mach stem MS$_1$, (a): $(\delta_1,\delta_2) = (54.5^{\circ},48.3^{\circ}$), (b): $(\delta_1,\delta_2) =~(84.5^{\circ},48.3^{\circ}$), (c): $(\delta_1,\delta_2) =~(77.5^{\circ},48.3^{\circ}$). It can be seen that when the curve satisfies the required condition $\delta_2 - \delta_1 \ll 1$ in radians, both curves are almost identical. Thus, the present curve represents a simpler form that can be used for modelling of curved shocks. }
\end{figure}

Here we illustrate the modification made to the second-order curve approximation made by \citet{Li1997}, as seen in Fig.~\ref{fig:libendor3}. The circular arc approximation in the osculating plane is used to model the curvature of the central Mach stem, for different slope angles. Little to no deviation occurs when the difference of the slopes $\delta = \delta_1 -\delta_2$ are small, which suggests that the osculating circle approximation is highly reliable and offers a simpler option for modelling curved planar shocks. Nevertheless, the computational results for the shape of the curved central Mach stem agree well with both the original and modified \citet{Li1997} curves (Fig.~\ref{fig:libendornumerical}), provided the slope angle $\delta_2$ is measured correctly.




 
%

\section{Conclusion}
The present analytical and numerical study investigated a peculiar shock reflection configuration occurring in non-uniform flow known as the cap-shock Mach reflection. We find that our theory successfully predicts the overall wave configuration of all geometrical features including the Mach stem height (both the circular and the curved one), its curvature, the shape of the slipstream and other characteristics and the flow-field properties within the subsonic pocket. It is demonstrated that our model shows excellent agreement with the numerical and analytical data in other, albeit limited, over-expanded nozzle studies that showed symmetric MR configurations; and agrees well with our numerical observation for the non-uniform csMR. 

In all cases, our model succesfully captures the von Neumann criterion for the oMR~$\leftrightarrow$~oRR transition, along with the asymmetric Mach stem height by including an averaged Mach number quantity at the centreline. It is found that despite the vastly different geometrical considerations used in wedge-induced shock reflections as compared to shock reflections inside of an over-expanded supersonic open jet, the analytical model still suffices to predict its Mach stem height in most cases. At higher shock angles and inflow Mach numbers, however, it appears that the change of the Mach stem height tends to be slightly lower than the wedge-induced case, as validated by the numerical data of \citet{hadjad2004} and analytical model of \citet{li1998mach}. This suggests the possibility that the MR configuration is more stable within the over-expanded supersonic jet as compared to the wedge-induced case.

\section{Acknowledgements}
The authors are grateful to the three anonymous referees who helped improve the manuscript. We also express gratitude toward Harald Kleine for helpful discussions. J.K.J.H.~acknowledges funding provided by the ANU Chancellor's International Scholarship. C.F.~acknowledges funding provided by the Australian Research Council (Future Fellowship FT180100495), and the Australia-Germany Joint Research Cooperation Scheme (UA-DAAD). We acknowledge high-performance computing resources provided by the National Computational Infrastructure (NCI) under the framework of the National Computational Merit Allocation Scheme (NCMAS), ANU Merit Allocation Scheme (ANUMAS) (grant~ek9), and the ANU Startup Scheme (grant~xx52), which is supported by the Australian Government. We further acknowledge high-performance computing resources provided by the Leibniz Rechenzentrum and the Gauss Centre for Supercomputing (grants~pr32lo, pr48pi and GCS Large-scale project~10391).

\section{Code and Data Availability}
The code for the analytic model and relevant datasets can be provided upon reasonable request to the author.



\section{Appendix A: Derivation for the shape of circular Mach stem height in the osculating plane with boundary conditions at its ends}\label{appendix:circle}

By \citet{Li1997}, a standard transformation between $(x,y) \rightarrow (x',y')$  and second order truncation of the Taylor series expansion yields a relation of the form:
\begin{equation}
    y' = f(x') = \frac{x'^2}{2x_2'}\varepsilon
\label{eqn:8.1}
\end{equation}
where the function describes a monotonic curve $Q_1Q_2$ with coordinates $(x_1,y_1)$, $(x_2,y_2)$ respectively, and where,
\begin{equation}
    x' = (x-x_1)\cos(\delta_1) + (y-y_1)\sin(\delta_1), \quad
    y' = -(x-x_1)\sin(\delta_1) + (y-y_1) \cos(\delta_1)
\end{equation}
 with $\varepsilon = \tan(\delta_2 - \delta_1)$ where $\delta_2$ and $\delta_1$ are the angles corresponding to the slopes of the points $Q_1$ and $Q_2$. The coordinates of the end points $\left[\left(x_{1}, y_{1}\right),\left(x_{2}, y_{2}\right)\right]$ also satisfy the relation:
$$
y_{2}-y_{1}=\tan \Lambda\left(\delta_{1}, \delta_{2}\right)\left(x_{2}-x_{1}\right)
$$
where $\Lambda$ is given by:
$$
\Lambda\left(\delta_{1}, \delta_{2}\right)=\arctan \left[\frac{2 \tan \delta_{1}+\tan \left(\delta_{2}-\delta_{1}\right)}{2-\tan \delta_{1} \tan \left(\delta_{2}-\delta_{1}\right)}\right]
$$

Now we proceed with the derivation by parameterising Eqn.~\ref{eqn:8.1}~in terms of the $(x',y')$ coordinate basis, which yields:
\begin{equation}
    \mathbf{r}(t) = \left\{t,\frac{t^2 \varepsilon }{2 x_2'}\right\}
\end{equation}
By differentiating twice and normalising,  the tangent and normal vector to the curves are obtained, which can subsequently be related to the flow-plane shock curvature \citep{molder},
\begin{equation}
   \mathbf{\hat{T}}(t) =  \left\{\frac{1}{\sqrt{\frac{t^2 \varepsilon ^2}{\left(x_2'\right)^2}+1}},\frac{t \epsilon }{x_2' \sqrt{\frac{t^2 \varepsilon ^2}{\left(x_2'\right)^2}+1}}\right\}
\end{equation}
\begin{equation}\label{eqn:normal}
\mathbf{\hat{N}}(t)  = \left\{-t \sqrt{\frac{t^2 \varepsilon ^2}{\left(x_2'\right)^2}+1} \sqrt{\frac{\varepsilon ^2 \left(x_2'\right)^2}{\left(t^2 \varepsilon ^2+\left(x_2'\right)^2\right)^2}},\frac{x_2' \sqrt{\frac{t^2 \varepsilon ^2}{\left(x_2'\right)^2}+1} \sqrt{\frac{\varepsilon ^2 \left(x_2'\right)^2}{\left(t^2 \varepsilon ^2+\left(x_2'\right)^2\right)^2}}}{\varepsilon }\right\}
\end{equation}
The principal shock curvature is then,
\begin{equation}\label{eqn:15.5}
        \mathcal{S}_a(t) = \frac{1}{|\mathbf{r}(t)}|\mathbf{\hat{T}}(t)| = \frac{\sqrt{\frac{\varepsilon ^2 \left(x_2'\right)^2}{\left(t^2 \varepsilon ^2+\left(x_2'\right)^2\right)^2}}}{\sqrt{\frac{t^2 \varepsilon ^2}{\left(x_2'\right)^2}+1}}
\end{equation}
This allows the radius of curvature, $ \rho(t) = 1/|\mathcal{S}_a(t)|$ to be evaluated. For the present derivation, we impose the condition that the circle in the osculating plane intersects the curve at exactly the midpoint of $x_1'$ and $x_2'$, yielding the point $(x_2'/2 , x_2'\varepsilon/8)$. Thus, the principal curvature at $t = x_2'/2$ is:
\begin{equation}
    \mathcal{S}_{a}(x_2'/2) = \frac{8 \sqrt{\frac{\varepsilon ^2}{\left(\varepsilon ^2+4\right)^2 \left(x_2'\right)^2}}}{\sqrt{\varepsilon ^2+4}}
\end{equation}
Using the normalised normal vector (Eqn \ref{eqn:normal}), the radius of curvature $ \rho(t = x_2'/2)$ and the intersecting point, the centre of the osculating circle $(z_1,z_2)$ can thus be found:
\begin{equation}
    (z_1,z_2) = \left\{x_2'/2, x_2'\varepsilon/8\right\} + \frac{1}{\mathcal{S}_{a}( x_2'/2)} \mathbf{\hat{N}}(x_2'/2) =  \left\{-\frac{1}{8} \varepsilon ^2 x_2',-\left(\frac{3 \varepsilon }{8}+\frac{1}{\varepsilon }\right) x_2'\right\}
\end{equation}
The circular arc in the osculating plane is therefore simply:
\begin{equation}
 (x' + \frac{1}{8} \varepsilon ^2 x_2')^2  + (y' -\left(\frac{3 \varepsilon }{8}+\frac{1}{\varepsilon }\right) x_2')^2 = \left(\frac{\sqrt{\varepsilon ^2+4}}{8 \sqrt{\frac{\varepsilon ^2}{\left(\varepsilon ^2+4\right)^2 \left(x_2'\right)^2}}}\right)^2
\end{equation}
Notice that this removes any constraint on $x_1'$, so that only knowledge of the parameter $x_2'$ is required for the prediction of Mach stem curvature. We also have $\delta_1 = \beta + \theta_1$ and $\delta_2 = \beta $ by approximate inspection of the slopes. 
\section{Appendix B: Geometrical relations for downstream flow properties}\label{appendix:geo}
Based on the nomenclature adopted within the constructed coordinate system, the location of the points $R$ and $T$ along the abcissa and ordinate can be readily obtained based on the geometrical configuration,
\begin{equation}\label{eq:18}
\left.\begin{array}{c}
 x_{Ru}=0, \quad y_{Ru} = 0, \quad x_{Ru} = 0, \quad y_{Rl} = 0,\\
x_{Tu} = H_u - H_{mu}, \quad y_{Tu} = x_{Tu} \cot{\beta_{u}}, \quad x_{Tl} = H_l - H_{ml}, \quad y_{T_l} = x_{Tl} \cot{\beta_{l}}\\
\end{array}\right\}
\end{equation}
We now proceed with the mathematical relations for the remaining points, refer to Fig. \ref{fig:crazy} for the following,
For the point $Q(x_Q,y_Q)$, 
\begin{equation}
    x_T\left[\cot(\beta) +\cot(\phi_2-\delta)\right] = x_Q \left[\cot(\delta) +\cot(\phi_2 - \delta)\right]
\end{equation}
where,
\begin{equation}
    y_Q = x_Q \cot(\delta)
\end{equation}
For the point $F(x_F,y_F)$,  
\begin{equation}
    x_T\left[\cot(\epsilon ) - \cot(\beta)\right] + x_Q\left[\cot(\delta) - \cot(\mu_C + \epsilon)\right]  = x_F \left[ \cot(\epsilon) - \cot(\mu_C + \epsilon)\right]
\end{equation}
where,
\begin{equation}
    y_F = \left[ x_F - x_T \right] \cot(\epsilon) + x_T \cot(\beta),\quad  \mu_C= \arcsin{\left(\frac{1}{M_C}\right)} 
\end{equation}
For the point $E(x_E,y_E)$:  
\begin{equation}
    x_E = H - H_s
\end{equation}
where,
\begin{equation}
    y_E = x_E\cot(\mu_E ) + x_Q\left[\cot(\delta) - \cot(\mu_E )\right],\quad
    \mu_E = \arcsin{\left( \frac{1}{M_E}\right)}
\end{equation}
For the point $J(x_J,y_J)$:
\begin{equation}
    x_J - x_E = \tan(\alpha )(y_J - y_E)
\end{equation}
where,
\begin{equation}\label{eqn:8.9}
x_J - x_F = \tan(\alpha - \epsilon)(y_J - y_F)
\end{equation}
This set of equations (\ref{eq:18}-\ref{eqn:8.9}), along with the $5$ equations in Sec. \ref{sec:2.4}  constitute the necessary conditions to solve for the Mach stem height $H_m$. Considering both the domains, we have a total of $28$ equations that are solved for $28$ unknowns. These unknowns are $x_Q,y_Q, x_T,y_T,x_E,y_E,x_J,y_J, M_C,M_E,H_m, H_s, \mu_E, \mu_J$, with an implicit subscript denoting upper and lower domains as usual. 
Following \citet{Roy2019}, we add another five equations where the Mach stem is required to have a continuous profile between the triple points of both the upper and lower domains, related through $\phi_4$, so that under a first-order approximation for point $G(x_G y_G)$, we have,
\begin{equation}
    y_G - y_T = 1/2\tan(\phi_4)H_m
\end{equation}
where,
\begin{equation}
 x_G = H
\end{equation}
and $H_u + H_l = H_T$, where $H_T$ is the total height of the domain. This yields a total of 33 equations altogether when the continuity boundary condition $y_{Gu} = y_{Gl}$ is imposed. These additional constraints therefore solve for $y_G, x_{Gu}, x_{Gl}, H_u, H_l$.

Also, since we do not utilise a universal coordinate system, the upper domain values will need to be shifted upwards. Therefore, once all values for the upper domain have been determined, the transformation $H_T - x_{ju}$, for each $j$ coordinate can then be applied, while keeping $y$ coordinates the same.
\section{Appendix C: Additional grid resolution studies and runs}\label{appendix:grid}
\begin{figure}
    \begin{subfigure}{0.48\textwidth}
    \centering
    \includegraphics[ width = \linewidth]{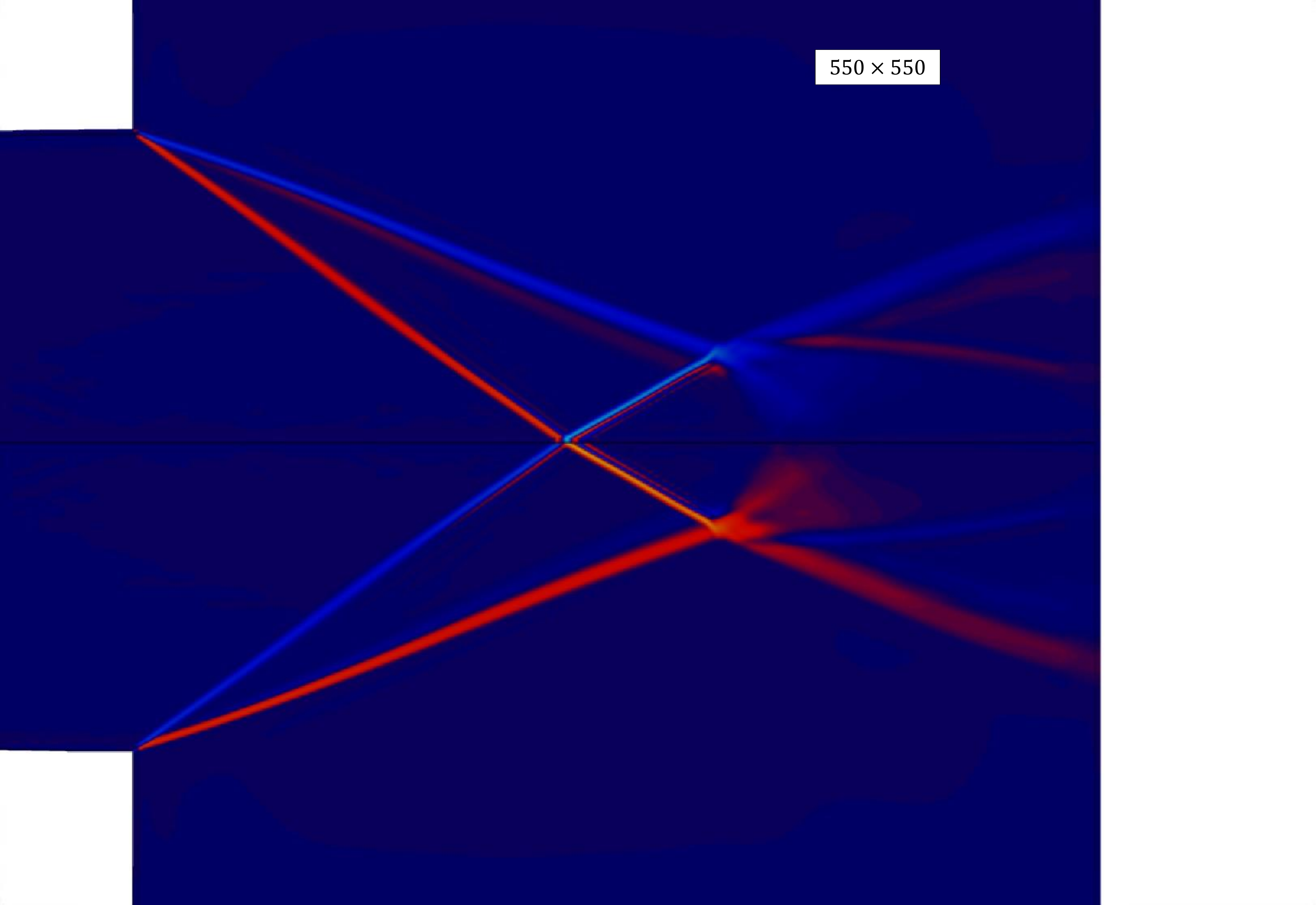}
    \caption{\label{fig:regref1}}
    \end{subfigure}
    \begin{subfigure}{0.48\textwidth}
    \centering
    \includegraphics[width = \linewidth]{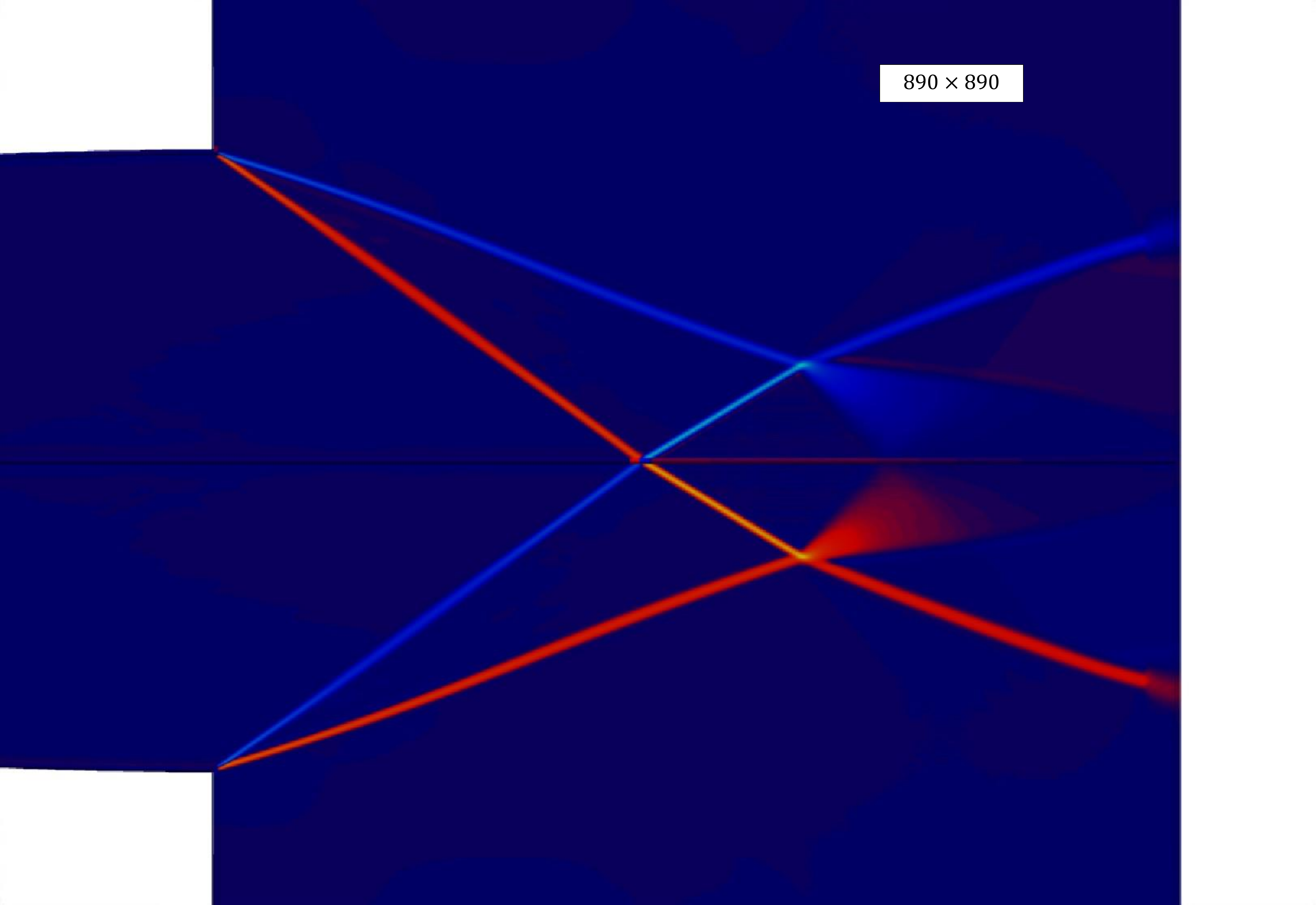}
    \caption{\label{fig:regref2}}
    \end{subfigure}
    
\caption{\label{fig:regref} Numerical schlieren images of regular reflection configuration outside an over-expanded jet with different grid densities: (a) $550 \times 550$ cells, (b) $890 \times 890$ cells}
\end{figure}

Additional grid resolution studies were done for a wide range of NPRs to investigate numerical convergence. Numerical schlieren images for the higher NPR range are shown in Fig. \ref{fig:regref1} - \ref{fig:regref2}. They illustrate the regular reflection configuration (see Fig 2.), otherwise known as two-shock system which consists of two incident shocks emanating from the nozzle lip, jet boundary interaction and subsequent shock-expansion-wave system. Both grids are already exceptionally dense and show good resolution, but the the finest one in Fig. \ref{fig:regref2} demonstrate grid convergence, where the shocks have been resolved within significantly more cells and so are much thinner. As mentioned earlier, the regular reflection configuration arises when the re-compression shock strength and turning angle between incident and reflected shocks can be matched according to the von Neumann criterion. 

\begin{figure}
 \centering
\includegraphics[width = 0.7\linewidth]{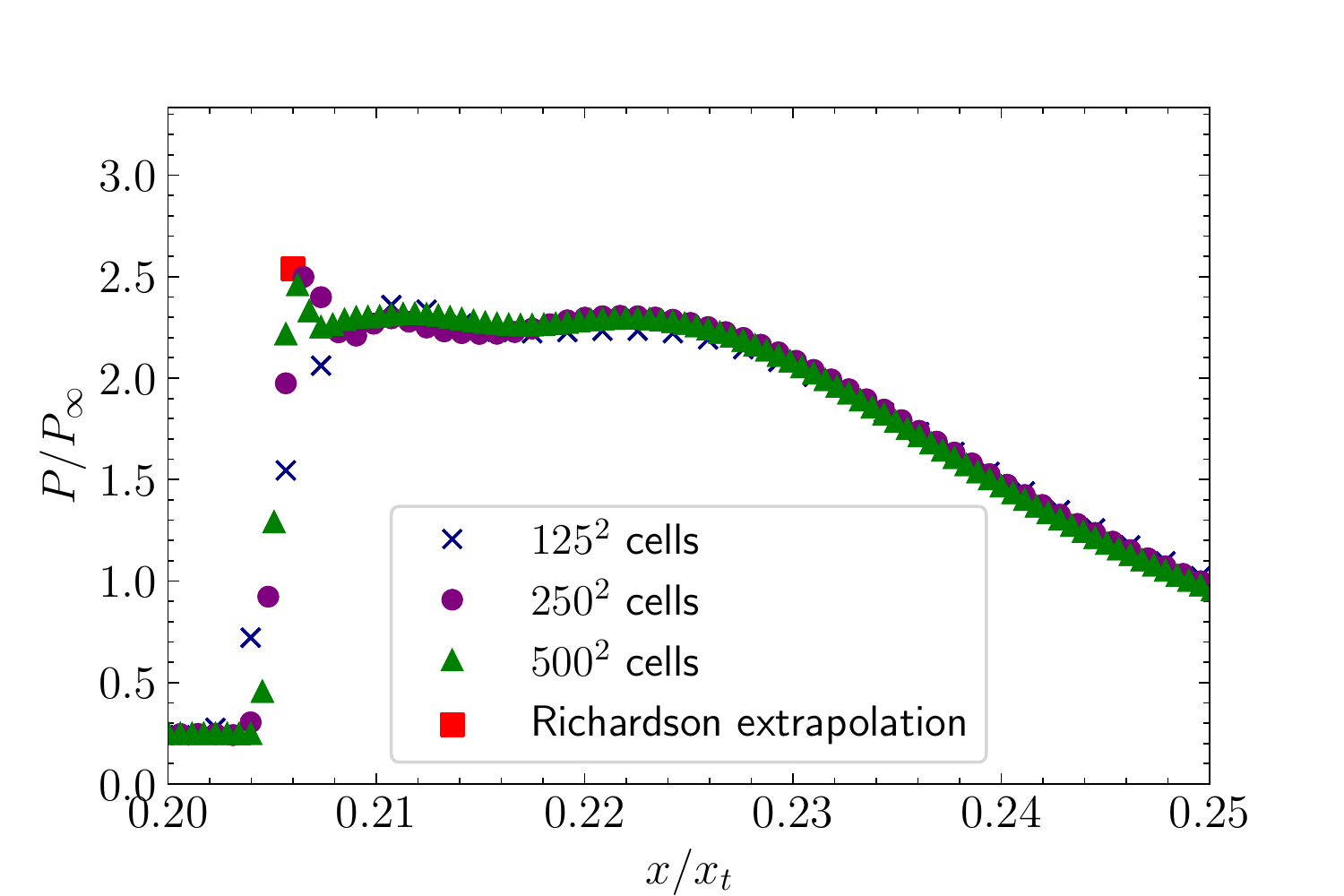}
  \caption{Non-dimensional static pressure plot along the centreline at $NPR = 100$, with increasing grid resolution ($125^2$, $250^2$ and $500^2$ cells), while keeping a constant refinement ratio. The maximum pressure value obtained via \citet{richardson1911ix} extrapolation is also shown. }
\label{fig:gridstudy}
\end{figure}

In addition, we examine quantitative measurements in the form of static pressure plots along the centreline with increasing grid resolution, as shown in Fig.~\ref{fig:gridstudy} for the NPR of 100 exhibiting a symmetric MR configuration. The values obtained thus ascertain convergence of the numerical solutions, as confirmed also by extrapolated data computed from the \citet{richardson1911ix} method. 
\bibliographystyle{jfm}
\bibliography{jfm-instructions}

\end{document}